\documentclass{article}

\usepackage[nonatbib,preprint]{neurips_2021}

\usepackage[utf8]{inputenc} 
\usepackage[T1]{fontenc}    
\usepackage{hyperref}       
\usepackage{url}            
\usepackage{booktabs}       
\usepackage{amsfonts}       
\usepackage{nicefrac}       
\usepackage{microtype}      
\usepackage{xcolor}         
\usepackage{graphicx}
\usepackage{amsmath}
\usepackage{amsthm}
\usepackage{caption}
\usepackage{subcaption}
\usepackage{enumitem}
\usepackage[numbers,sort&compress]{natbib}
\usepackage{rotating} 

\newcommand{\BE}{{\mathbb{E}}}	
\newcommand{\BF}{{\mathbb{F}}}	
\newcommand{\BI}{{\mathbb{I}}}	
\newcommand{\BP}{{\mathbb{P}}}	
\newcommand{\BR}{{\mathbb{R}}}	
\newcommand{\BV}{{\mathbb{V}}}	
\newcommand{\raro}{\rightarrow}	
\newcommand{\cA}{\mathcal{A}}	
\newcommand{\cI}{\mathcal{I}}	
\newcommand{\cX}{\mathcal{X}}	
\newcommand{\trace}{\mathrm{trace}}

\def\baq#1\eaq{\begin{align}#1\end{align}}
\def\ban#1\ean{\begin{align*}#1\end{align*}}
\newtheorem{Theorem}{Theorem}
\newtheorem{corollary}{Corollary}

  \newtheoremstyle{Example}{\topsep}{\topsep}%
     {}
     {}
     {\bfseries}
     {:}
    {0.9mm}
     {\thmname{#1}\thmnumber{ #2}\thmnote{(\it #3)}}
		
  \theoremstyle{Example}
  \newtheorem{Example}{Example}[section]

\title{Generalized Multivariate Signs for Nonparametric Hypothesis Testing
in  High Dimensions}

\author{%
  Subhabrata Majumdar\\
  Data Science and AI Research\\
  AT\&T Chief Data Office\\
  New York, NY 10007\\
  \texttt{subho@att.com}
  \And
  Snigdhansu Chatterjee\\
  School of Statistics\\
  University of Minnesota Twin Cities\\
  Minneapolis, MN 55455\\
  \texttt{chatt019@umn.edu}
}

\begin{document}

\maketitle

\begin{abstract}
High-dimensional data, where the dimension of the feature space is much larger than sample size, arise in a number of statistical applications. In this context, we construct the generalized multivariate sign transformation, defined as a vector divided by its norm. For different choices of the norm function, the resulting transformed vector adapts to certain geometrical features of the data distribution. Building up on this idea, we obtain one-sample and two-sample testing procedures for mean vectors of high-dimensional data using these generalized sign vectors. These tests are based on U-statistics using kernel inner products, do not require prohibitive assumptions, and are amenable to a fast randomization-based implementation. Through experiments in a number of data settings, we show that tests using generalized signs display higher power than existing tests, while maintaining nominal type-I error rates. Finally, we provide example applications on the MNIST and Minnesota Twin Studies genomic data.
\end{abstract}

\section{Introduction}
\label{Sec:Introduction}

In many scientific problems, most features or variables generally have no relation with the label or feature under study, and sparsity is the dominant paradigm. However, scientific tests suitable for composite hypotheses to set several parameter values to zero are lacking in many respects. For example, in studies linking genetic information to cancer or diabetes, single nucleotide polymorphisms (SNPs) are often considered one at a time, leading to multiple testing issues and inefficiencies \cite{ManolioEtal09}. 

Classical joint hypothesis tests on $p$ parameters, for example using Hotelling's $T^{2}$ test statistic \citep{ref:Anderson_Multivariate_Book, ref:Muirhead_Multivariate_Book}, are inadequate in high dimensions, i.e. when $p$ is very large (in the magnitudes of hundreds or thousands). An associated problem for hypothesis testing in high-dimensional real data is that there are often outliers, inliers and other rogue observations in the dataset, and most multivariate data science methods, such as principal component analysis and its kernel and nonlinear variants are sensitive to such aberrant observations: hence there is a need for robust test statistics 
\citep{ref:JASA151658_HDTest_Lan}. To add to the above issues, we reveal later in this paper that most proposed hypothesis test statistics perform poorly and have low power under at least one of the possible alternative scenarios where the data contains either a very sparse signal, or a moderate sized sparse signal, or low but non-sparse signal. 

\paragraph{Related work}
Attempts have been made in the past few years to extend the paradigm of joint hypothesis tests on $p$ parameters, where $p$ is large. Important references in this regard are 
\cite{ref:StatSinica96311_HDTest_BaiSaranadasa, ref:JMVA08386_HDTest_SrivastavaDu, ref:JMVA09518_HDTest_Srivastava, ref:AoS10808_HDTest_ChenQin, 
ref:JRSSB14349_HDTest_Cai, ref:JASA151658_HDTest_Lan}. 
Assume that the data 
consists of observations $\{ X_{i} \in \BR^{p}, \ i = 1, \ldots, n \}$, and each $X_{i}$ has a distribution centered at $\mu \in \BR^{p}$. The null hypothesis of interest is $H_{0}: \mu = \mu_{0}$ for a pre-specified $\mu_{0} \in \BR^{p}$, while the alternative hypothesis may be either that ($i$) one (or a very small) number of elements of $\mu$ do not match the corresponding elements of $\mu_{0}$
(very sparse alternative), or ($ii$) a small fraction of the elements of $\mu$ do not match the corresponding elements of $\mu_{0}$ (sparse alternative), or
($iii$)  several elements of $\mu$ do not match the corresponding elements of $\mu_{0}$ (dense alternative). A two-sample version of the above problem can also be defined similarly.

The above listed references assume that the observations $X_{1}, \ldots, X_{n}$ are independent, identically distributed (iid) with an absolutely continuous distribution,  and a substantial number of these further specify that this common distribution is Gaussian, sometimes with a diagonal covariance matrix. Under such restrictive conditions, building on the previous series or papers, the state-of-the-art in high-dimensional hypothesis testing is given in 
\cite{ref:JASA151658_HDTest_Lan}. Their proposed test statistic is
$$T_{n} = \sum_{i \not= j} {\frac {X_{i}^{T}X_{j}}{|X_{i}|_{2} |X_{j}|_{2}}},$$
where $| \cdot |_{2}$ is the Euclidean norm. The two-sample version of this was later developed  in \cite{ref:JASA16721_HDTest_Changliang, ref:AoS17771_HDTest_Probal}. 

\paragraph{Our contributions}
In this paper, we generalize the above line of work through several major extensions:
\begin{itemize}[leftmargin=*]
\setlength\itemsep{0em}
\item For a general scaling function $D: \BR^{p} \raro [0, \infty)$, define the \textit{generalized sign} for $x \in \BR^{p}$ as $s = x/ D (x). \cI_{ \{ D (x) \not = 0\}}$. Examples of scaling functions are $\ell_{k}$-norms 
$|x|_{k} = ( \sum_{j = 1}^{p} | x_{j} |^{k} )^{1/k}$, and data depth functions 
\cite{ref:AoS99783_Depth_LiuSingh, ref:AoS00461_ZuoSerfling, ref:DIMACS061_Serfling}. 
We propose using (essentially) the test statistic 
$$ T_{n} = {\binom{n}{2}}^{-1} \sum_{1 \leq i < j \leq n} S_{i}^{T} S_{j},$$
where  $S_{i} = X_{i}/D (X_{i}). \cI_{ \{ D (X_{i}) \not = 0\}}$. This is a direct extension of \cite{ref:JASA151658_HDTest_Lan}, which used the $\ell_{2}$-norm as the scaling function. Even though we obtain parallel results, we utilize  completely different theoretical machinery that do not require a number of restrictive conditions assumed in recent papers on related topics. The choice of different norms for the scaling function has interesting consequences for the power and efficiency of the test. We sometimes find $\ell_{1}$ or $\ell_{\infty}$ norm as better choices compared to $\ell_{2}$ norm, as demonstrated in our numerical experiments.

\item Next, for a suitable kernel function $K : \BR^{p} \times \BR^{p} \raro \BR$, we define the new test statistic 
$$ T_{K, n} = {\binom{n}{2}}^{-1} \sum_{1 \leq i < j \leq n} 
K \bigl( S_{i},  S_{j} \bigr) - K \bigl(0, 0 \bigr).$$
Using the kernel trick in place of Euclidean inner product results in significant improvement in the power of the test under different alternatives. We present theoretical results on selecting kernels to optimize power for one-sample and two-sample testing problems.

\item We illustrate the competitive efficiency of the proposed methods in several simulated data experiments, and in two real data examples. For the widely-used MNIST dataset, our test is able to distinguish between ambiguous numbers such as 9 and 4, or 5 and 8. For the challenging second example, concerned with detecting SNP-level weak signals associated with quantitative traits such as alcoholism and drug abuse, we identify several novel associations between traits and genes, and corroborate them with existing domain knowledge.
\end{itemize}



Since there are considerable technicalities in this paper, we first present in Section~\ref{Sec:Euclidean} results for the test statistic involving the Euclidean inner product. Following this, in Section~\ref{Sec:Kernel} results for a general kernel are presented. Due to limited space, we relegate all proofs and results from several numerical experiments to the supplementary Appendix.

\section{Euclidean Test Statistic}
\label{Sec:Euclidean}

We use the notation $| \cdot |_{k}$ to denote the $\ell_{k}$ norm, except that for simplicity we also use $| \cdot |$ (without suffix) for the $\ell_{2}$ (Euclidean norm.) Our examples will typically involve $k = 1, 2, \infty$, that is, we will primarily concern ourselves with the $\ell_{1}, \ell_{2}$ and $\ell_{\infty}$ norms. For a square symmetric matrix $A$, we denote by $A >0$ the fact that $A$ is positive-definite.

\paragraph{Formulation}
Suppose $X_{i}$'s are  iid copies of some absolutely continuous random variable $X \in \cX \subseteq \BR^{p}$ such that
\baq 
\BE X_{1} & = \mu \in \BR^{p},  \label{eq:mu}\\
\BV X_{1} & = \Sigma > 0,  \label{eq:Sigma}\\
\BE | X_{1} |^{8} & = O (p) < \infty.  \label{eq:XMoment}
\eaq
In view of the above, we can write $X_{i} = \mu + \Sigma^{1/2} Z_{i}$,  where $Z_{i}$'s are iid, with
$$ \BE Z_{i} = 0 \in \BR^{p}, \quad
\BV Z_{i} = \BI_{p}, \quad
\BE |Z_{i}|^{8} = O (p) < \infty. $$
We assume that that the eigenvalues of $\Sigma$ are bounded above and below
\baq 
\lambda_{j} (\Sigma) \in (C^{-1}, C) \ \ \text{ for some } \ C > 0, 
\text{ for all } j = 1, \ldots, p. 
\label{eq:Eigenvalue}
\eaq
Also, we will use the notation $C$, with or without subscripts, as generic for constants below, and these do not depend on $n$ or $p$.

Define a function $D: \cX \raro \BR_{+}$ with the property that for a sufficiently small 
$\epsilon > 0$ we have 
\baq
\bigl\{x: D (x) \leq \epsilon \bigr\}
\subseteq 
\bigl\{x: |x|_{1} \leq C_{0} \epsilon \bigr\}, 
\label{eq:DCondition}
\eaq
for some $C_{0} > 0$. Standard examples of the $D (\cdot)$ function are $\ell_{k}$ norms for 
$k = 1, 2, \infty$ for which $C_{0} = 1$ suffices. We can also use $
D (x) = ( \tilde{D} (x; \BF_{0}) )^{-1} |x|_{k}
$, where $\tilde{D} (\cdot; \BF_{0})$ is a \textit{data depth} with respect to some distribution $\BF_{0}$ centered at $0 \in \BR^{p}$. Using the standard assumption that 
$0 < \tilde{D} (x; \BF_{0}) \leq 1$ for all $x \in X$ without loss of generality, the choice of $C_{0} = 1$ again suffices. 
Note that the use of the $| \cdot |_{1}$-norm ball in 
\eqref{eq:DCondition} is merely for convenience, we can easily use some other norm, if necessary by scaling $\epsilon$ appropriately. 

We assume that the density $f (\cdot)$ of $X$ satisfies the following condition:
There exists a $\delta > 0$ and a positive constant $C_{1} > 0$ such that
\baq
f (x) \leq C_{1}, \ \ \text{ for all } |x|_{1} < \delta. 
\label{eq:Density}
\eaq
This condition merely states that the density of $X$ is bounded in a neighborhood of 
$0 \in \BR^{p}$. Without loss of generality, we can select $\epsilon$ from 
\eqref{eq:DCondition} and $\delta$ from \eqref{eq:Density} to satisfy 
$\epsilon \leq \delta$. 

For a decreasing sequence $\{ \epsilon_{n} \}$ to be specified later, we define the 
\baq 
S_{i} = {\frac {X_{i}}{D (X_{i})}} \cI_{\{ D (X_{i}) > \epsilon_{n} \}}.
\label{eq:S} 
\eaq 
Note that when $D (x) = |x|$, this is the well-known \textit{spatial sign} vector corresponding to $X_{i}$ \cite{SpatialSign}. Here in \eqref{eq:S}, we adopt a much more broad notion of (weighted and generalized) sign. 

\paragraph{Generalized sign test}
Using the generalized spatial sign defined above, we define our test statistic is defined as a one-sample U-statistic:
\baq 
T_{D, n} = {\binom{n}{2}}^{-1} \sum_{1 \leq i < j \leq n} S_{i}^{T} S_{j}. 
\label{eq:T_E}
\eaq

We now obtain the asymptotic distribution of $T_{E, n}$, under the conditions \eqref{eq:mu}--\eqref{eq:Eigenvalue} stated above, and a minor moment condition
\baq 
\BE \bigl( D (X_{i}) - \BE D (X_{i}) \bigr)^{8} | X_{i}|^{8} = O (p). 
\label{eq:DZ1}
\eaq

\begin{Theorem}
\label{Thm:Euclidean}
Let  $\Delta = \BE D (X_{1})$. Under the conditions given above, 
$ n \Delta^{2} p^{-1} T_{D, n} $ has the asymptotic distribution of a degenerate $U$-statistic, with a non-centrality parameter of 
$\nu = n p^{-1} | \mu|^{2}$. 
Thus, the non-centrality parameter vanishes under the null hypothesis $H_0: \mu=0$, with limiting distribution
\ban
T_{D,n} \leadsto \sum_{k=1}^\infty \lambda_k (W_k -1),
\ean 
where $\{ W_k\}$ are iid $\chi_1^2$ random variables and $\{\lambda_k\}$ are the (non-zero) eigenvalues of a certain operator based on the distribution of $X$ 
.
\end{Theorem}
The proof of the above result uses standard tools from U-statistics asymptotics \citep[Chapter 5]{serflingbook} and Hilbert algebra \cite{mmd}. We provide it in the Appendix.






\section{Kernel tests based on the generalized sign}
\label{Sec:Kernel}

\subsection{One-sample test}
We now introduce the kernel version of $T_{E, n}$, for a suitable kernel $K (\cdot, \cdot)$:
\baq 
T_{K, n} = {\binom{n}{2}}^{-1} \sum_{1 \leq i < j \leq n} 
K \bigl( S_{i},  S_{j} \bigr) - K \bigl(0, 0 \bigr). 
\label{eq:T_K}
\eaq
Note that $K$ is a symmetric function in its arguments, that is $K (x, y) = K (y, x)$ 
for all $x, y \in \cX$.
We assume that the kernel $K (\cdot, \cdot)$ is twice continuously differentiable.
We define the functions
$K_{1} (x, y) = {\frac{\partial}{\partial x}} K(x, y)$, 
$K_{2} (x, y) = {\frac{\partial^{2}}{\partial x^{2}}} K(x, y)$, 
$K_{11} (x, y) = {\frac{\partial}{\partial y}} {\frac{\partial}{\partial x}} K(x, y)$. 

We are now  state our main theorem involving kernel functions. It contains several cases, to quantify the different ways kernel functions may be used for high-dimensional hypothesis testing.

\begin{Theorem} 
\label{Theorem:KernelCLT}
Consider a kernel function satisfying the technical conditions stated above, and the one-sample of problem of testing $H_0: \mu=\mu_0$ vs. $H_a: \mu \neq \mu_0$. Let $\Delta = O (a_{p})$, and define the sets
\begin{align*}
\mathcal E_1 &= \left\{ \mu \in \mathbb R^p:
a_{p}n^{1/2} [ K (\Delta^{-1} \mu, \Delta^{-1} \mu) - K (0, 0) ] := \delta_1 < \infty \right\},\\
\mathcal E_2 &= \left\{ \mu \in \mathbb R^p:
a_{p}^2 n [ K (\Delta^{-1} \mu, \Delta^{-1} \mu) - K (0, 0) ] := \delta_2 < \infty \right\}.
\end{align*}

\begin{description}[leftmargin=*]
\item ($i$) Assume that 
$K_{1} (\nu, \nu) \not= 0 \in \BR^{p}$ for any $\nu \in \BR^{p}$. Then 
under $H_0$, $a_{p}n^{1/2} T_{K, n}$ has an asymptotic distribution 
$N \bigl( 0, K_{1}^{T} (0, 0) \Sigma K_{1} (0, 0) \bigr)$. 
Further, when $\mu \in \mathcal E_1$, $a_{p}n^{1/2} T_{K, n}$ has asymptotic distribution $N \bigl( \delta_{1}, K_{1}^{T} (\Delta^{-1} \mu, \Delta^{-1} \mu) \Sigma K_{1} (\Delta^{-1} \mu, \Delta^{-1} \mu) \bigr)$ under $H_a$.

\item ($ii$)  Assume that 
$K_{1} (\nu, \nu) \not= 0 \in \BR^{p}$ for any 
$\nu \in \BR^{p}\setminus \{ 0 \}$, but $K ( 0, 0) = 0 \in \BR^{p}$. Then 
under $H_0$, $a_{p}^{2} n T_{K, n}$ has a non-Gaussian asymptotic distribution  involving unknown parameters. When $\mu \in \mathcal E_2$, $a_{p}^{2} n T_{K, n}$ has the asymptotic distribution 
$N \bigl( \delta_{2}, K_{1}^{T} (\Delta^{-1} \mu, \Delta^{-1} \mu) \Sigma K_{1} (\Delta^{-1} \mu, \Delta^{-1} \mu) \bigr)$ under $H_a$. 

\item ($iii$)  Assume that 
$K_{1} (\nu, \nu) = 0 \in \BR^{p}$ for any 
$\nu \in \BR^{p}\setminus \{ 0 \}$, but $K ( 0, 0) \not= 0 \in \BR^{p}$. Then 
under $H_0$, $a_{p}n^{1/2} T_{K, n}$ has an asymptotic distribution 
$N \bigl( 0, K_{1}^{T} (0, 0) \Sigma K_{1} (0, 0) \bigr)$. 
When $\mu \in \mathcal E_1$, $a_{p}n^{1/2} T_{K, n} -\delta_{1}$ has 
a non-Gaussian asymptotic distribution with unknown parameters
under $H_a$. 

\item ($iv$)  Assume that 
$K_{1} (\nu, \nu) = 0 \in \BR^{p}$ for any 
$\nu \in \BR^{p}$. Then under $H_0$, $a_{p}^{2} n T_{K, n}$ has a non-Gaussian asymptotic distribution  involving unknown parameters. When $\mu \in \mathcal E_2$, $a_{p}^{2} n T_{K, n} - \delta_{2}$ has a non-Gaussian asymptotic distribution  involving unknown parameters under $H_a$.  
\end{description}

\end{Theorem}

The non-Gaussian distributions in Theorem~\ref{Theorem:KernelCLT} are structurally all similar and relate to the limiting distribution of degenerate $U$-statistics; for example see \citep[Chapter 5]{serflingbook} and Theorem~\ref{Thm:Euclidean}. However, their parameters differ based on the conditions. Also, note that under the alternative for some values of $\mu$, it is possible that $\delta_{1}$ or $\delta_{2}$ tends to infinity as $n \raro \infty$, as is the case often for non-contiguous alternatives in traditional hypothesis testing. These cases are trivially covered above, where 
under the alternative the limiting distribution tends to be degenerate. 
Interestingly, Theorem~\ref{Theorem:KernelCLT} also allows us to consider the more nuanced case of contiguous alternatives where $|\delta| \raro 0$ as $n \raro \infty$. 


Using Theorem~\ref{Theorem:KernelCLT}, we are in a position to consider different alternatives. We can group these alternatives using the rate condition 
$ |\mu |^{2} = O  \bigl( c_{p} b_{n}^{2}  \bigr)$, where $c_{p} \in \{ 1, \ldots, p \}$
denotes a level of sparsity, and $b_{n}$ controls the size of deviance from zero of any specific element of $\mu$. Interesting special cases for which optimal power functions can be obtained using Theorem~\ref{Theorem:KernelCLT} include ($a$) $c_{p} = O(1)$ and $b_{n} = O(1)$ 
and hence $ |\mu |^{2} = O  \bigl( c_{p} b_{n}^{2}  \bigr) = O (1) $ when
$\mu$ is sparse with a few prominent non-zero elements, ($b$) $c_{p} = O ( p)$ and 
$b_{n} = O ( n^{-1/2} p^{-1/2})$  and hence $ |\mu |^{2} = O  \bigl( c_{p} b_{n}^{2}  \bigr) = O (n{-1}) $ when most elements of $\mu$ are non zero but small in size. A particularly challenging case is  ($c$) 
$b_{n} = O ( n^{-1/2} p^{-1/2})$ and $c_{p} = O(1)$, 
which arises when $\mu$ is sparse with a few small entries, and 
 $ |\mu |^{2} = O  \bigl( n^{-1} p^{-1}  \bigr) = O (1) $. Other cases where $\mu$ has intermediate levels of sparsity and signal strength are also easily covered by Theorem~\ref{Theorem:KernelCLT}.

Let us explore some examples below:
\begin{Example}[Linear kernel]

Here we have $K (x, y) = x^{T} y$ for $x, y \in \BR^{p}$. Then we have 
$K_{1} (x, y) = {\frac{\partial}{\partial x}} K(x, y) = y \in \BR^{p}$, 
$K_{2} (x, y) = {\frac{\partial^{2}}{\partial x^{2}}} K(x, y) = 0 \in \BR^{p \times p}$, 
$K_{11} (x, y) = {\frac{\partial}{\partial y}} {\frac{\partial}{\partial x}} K(x, y) 
= \BI \in \BR^{p \times p}$. Thus in this case the appropriate scale is 
$a_{p}^{2} n$, and this pertains to case ($ii$) of Theorem~\ref{Theorem:KernelCLT}. 

Note that $K (\nu, \nu) = | \nu |_{2}^{2}$, and thus $K (0, 0) = 0$ and 
$K (\Delta^{-1} \mu, \Delta^{-1} \mu) = \Delta^{-2} | \mu|^{2} \not= 0$ for 
$\mu \not= 0 \in \BR^{p}$. Thus we have 
$\delta = a_{p}^{2} n \bigl( K (\Delta^{-1} \mu, \Delta^{-1} \mu) - K (0, 0) \bigr)
= a_{p}^{2} n \Delta^{-2} | \mu|^{2} = O \bigl( n | \mu|^{2} \bigr)$. 

Note that these details match with our explicit computations from the previous section.
\end{Example}

\begin{Example}[Polynomial kernel]
\label{ex:expoly}
Here we have $K (x, y) = \bigl( x^{T} y + a \bigr)^{b}$ for $x, y \in \BR^{p}$. 
Then we have 
$K_{1} (x, y) = {\frac{\partial}{\partial x}} K(x, y) 
=  b \bigl( x^{T} y + a \bigr)^{b -1} y \in \BR^{p}$, 
$K_{2} (x, y) = {\frac{\partial^{2}}{\partial x^{2}}} K(x, y) = 
b (b - 1)\bigl( x^{T} y + a \bigr)^{b -2} yy^{T} \in \BR^{p \times p}$, 
$K_{11} (x, y) = {\frac{\partial}{\partial y}} {\frac{\partial}{\partial x}} K(x, y) 
= b (b - 1)\bigl( x^{T} y + a \bigr)^{b -2} yy^{T} +  
b \bigl( x^{T} y + a \bigr)^{b -1} \BI \in \BR^{p \times p}$. Thus in this case the appropriate scale is 
$a_{p}^{2} n$, and this pertains to case ($ii$) of Theorem~\ref{Theorem:KernelCLT}. 
However, the null distribution is different from that of the case of the linear kernel.

Note that we have
$\delta = a_{p}^{2} n \bigl( K (\Delta^{-1} \mu, \Delta^{-1} \mu) - K (0, 0) \bigr)
\approx a_{p}^{2} n \Delta^{-2}  b a^{b - 1} | \mu|^{2} = 
O \bigl( b a^{b - 1} n | \mu|^{2} \bigr)$. With $b = O(p)$ and $a \raro \infty$
using the polynomial kernel we can have $| \delta | \raro \infty $ even under the 
sparse, weak signal condition encapsulated in ($c$) above. 

\end{Example}

It is interesting to note that the Gaussian or radial basis kernel is not very suitable for our specific hypothesis testing problem. This is because for the one sample problem $H_0: \mu=\mu_0$, taking pairwise differences used in calculating the Gaussian kernel eliminates any first-order effect of the mean parameter in the difference $\Delta$.

\subsection{The two-sample problem}
Consider now iid copies of two absolutely continuous random variable $Y_1, Y_2 \in \cX \subseteq \BR^{p}$ such that
\baq 
\BE Y_{1} = \mu_1 \in \BR^{p},  Y_{2} = \mu_2 \in \BR^{p} \label{eq:mu2}\\
\BV Y_{1} = \BV Y_{2} = \Sigma > 0,  \label{eq:Sigma2}\\
\BE | Y_{1} |^{8} = O (p) < \infty, \BE | Y_{2} |^{8} = O (p) < \infty  \label{eq:XMoment2}
\eaq
We also assume that the bounded eigenvalue condition \eqref{eq:Eigenvalue} continues to hold for $\Sigma$. Under this setup, consider the two-sample mean testing problem $H_0^2: \mu_1 = \mu_2$ vs. $H_a^2: \mu_1 \neq \mu_2$. We tackle this problem using a two-sample version of the kernel test statistic \eqref{eq:T_K}, based on a size $n$-sample composed of $n_1$ and $n_2$ samples of $Y_1, Y_2$, respectively:
\baq 
T_{2,K,n_1,n_2} &= \frac{2}{n_1(n_1-1)} \sum_{1 \leq i < i' \leq n_1} 
\tilde K \bigl( S_{1i},  S_{1i'} \bigr) +
\frac{2}{n_2(n_2-1)} \sum_{1 \leq j < j' \leq n_2} 
\tilde K \bigl( S_{2j},  S_{2j'} \bigr) \notag\\
&- \frac{2}{n_1 n_2} \sum_{1 \leq i \leq n_1, 1 \leq j \leq n_2} 
\tilde K \bigl( S_{1i},  S_{2j'} \bigr), \label{eq:T_2K}
\eaq
where $\tilde K(\cdot,\cdot) = K(\cdot,\cdot) - K(0, 0)$ is the centered kernel. The following now holds, in presence of mild assumptions on the comparative values of sample sizes $n_1, n_2$:

\begin{Theorem}\label{thm:twosample}
Suppose that as $n_1, n_2 \rightarrow \infty$, we have $n_1/n \rightarrow \rho_1, n_1/n \rightarrow \rho_2 = 1 - \rho_1$. Also, define
\begin{align*}
& \mathcal E_{12} = \left\{ (\mu_1, \mu_2) \in \mathbb R^p \times \mathbb R^p: \right.\\
& \left. a_{p}n^{1/2} [ K (\Delta^{-1} \mu_1, \Delta^{-1} \mu_1) +
K (\Delta^{-1} \mu_2, \Delta^{-1} \mu_2) ] -
2 K (\Delta^{-1} \mu_1, \Delta^{-1} \mu_2) ] := \delta_{12} < \infty \right\}.
\end{align*}
Then, under $H_0^2$,
$$ a_p^2 n T_{2,K,n_1,n_2} \leadsto \sum_{k=1}^\infty \gamma_k (A_k-1),$$
where $A_k$ are iid $\chi_1^2$ random variables, and $\gamma_k$ are the eigenvalues of a certain operator depending on the distributions of $Y_1$ and $Y_2$. Under the alternative $H_a^2$, if $(\mu_1, \mu_2) \in \mathcal E_{12}$, then $a_p n^{1/2} T_{2,K,n_1,n_2}$ is asymptotically Gaussian with mean $\delta_{12}$ and finite variance.
\end{Theorem}
A consistency result against families of local alternatives follows immediately.
\begin{corollary}
Under the sequence of contiguous alternatives $H_{an}^2$ such as $n | \delta_{12} |^2 \rightarrow \infty $, for any $\alpha \in (0,1)$ the level-$\alpha$ test based on $T_{2,K,n_1,n_2}$ is consistent, i.e. as $n \rightarrow \infty$ the power approaches 1. 
\end{corollary}
While the form of asymptotic distributions in Theorem~\ref{thm:twosample} is similar to the well-known result on kernel two-sample tests by Gretton et al \cite{mmd}, we note that our null hypothesis of mean equality is weaker, and requires a very different proof.

\subsection{Computation}
As we have seen in theorems \ref{Thm:Euclidean}, \ref{Theorem:KernelCLT} and \ref{thm:twosample}, null distributions of test statistics based on generalized signs often involve non-Gaussian distributions and unknown parameters. In such situations, randomization methods such as bootstrap and permutation are popular in numerically estimating the null distribution. However, in a one-sample U-statistic setting, bootstrap equivalents of the test statistic $T_{K,n}$ have a different variance than that of $T_{K,n}$ under null. Wang and Xu \cite{WANG2019160} showed this for the untransformed case: $D(X_1) = 1$, and the generalization is straightforward.

We use a multiplier randomization method instead to approximate the one-sample null distribution. Consider $\alpha_1, \ldots, \alpha_n$ to be i.i.d. Rademacher random variables: $P(\alpha_1 = 1) = P(\alpha_1 = -1) = 0.5$.
Conditioning on our original sample $(X_1, \ldots, X_n)$, we obtain samples from the null distribution based on $(\alpha_1 X_1, \ldots, \alpha_n X_n)$:
$$ T^\alpha_{K, n} = 
{\binom{n}{2}}^{-1} \sum_{1 \leq i < j \leq n} 
K \bigl( S^\alpha_{i},  S^\alpha_{j} \bigr) - K \bigl(0, 0 \bigr),$$
with $S^\alpha_i, 1 \leq i \leq n$ being the generalized signs constructed using $\alpha_i X_i$. We repeat this $B$ times for a large number of $B$, using multiple samples of the sign-flips $(\alpha^*_1, \ldots, \alpha^*_n)$, to compute $T^*_1, \ldots, T^*_B$. Thus the approximate $p$-value for our test is
$$ \hat p = \frac{1}{B+1} \left( 1 + \sum_{b=1}^B \mathcal I_{T^*_b \geq T_{K,n}}\right). $$
Consequently we accept $H_0: \mu=0$ at confidence level $\alpha \in (0,1)$ if $\hat p \leq \alpha$, reject otherwise.

For the two-sample procedure, a conventional permutation test suffices. We draw without-replacement samples of size $n_1$ from the combined sample, label them as group 1, label the rest of them as group 2, and compute the two-sample test statistic from this resample using \eqref{eq:T_2K}. A permutation $p$-value can be computed similarly as above using the size-$B$ resamples.

Implementing either of the above procedures for our kernel choices ---linear and polynomial---requires minimal computation beyond what is required originally for $T_{K,n}$ or $T_{2,K,n_1,n_2}$. The inner-products $S_i^T S_j, 1 \leq i \leq j \leq n$ can be computed once and stored. This takes $O(n^2 p)$ time. Following this, computing each copy of $T^*_b$ only involves multiplying $S_i^T S_j$ with $\alpha^*_i \alpha^*_j$, which takes $O(n^2)$ time to iterate over all sample pairs. Thus the overall time complexity to generate the null distribution is $O(n^2 p + n^2 B)$. Since the choice of $B$ is fixed, when $p$ is large this complexity is $O(n^2 p)$.










\section{Experimental evaluation}
\label{sec:sim}
We now evaluate the performance of our proposed tests on a number of data settings, values of the kernel parameters, and using true vs. original null distribution $p$-values. We also present comparison with several existing methods. We use type-I and $1-$type-II errors of the tests, conventionally known as nominal size and power, for evaluation. At different amounts of departure from the null hypotheses, we average the proportion of rejections for a test at confidence level $\alpha=0.05$ over 1000 independent datasets to obtain the power. The same rejection proportion calculated at the null hypothesis gives the empirical size, and ensures the tests are well-calibrated. To generate approximate null distribution using the randomization procedure, we use $B=1000$.

\paragraph{One-sample tests}
We consider $n=100$ samples drawn from a $p=300$-dimensional probability distribution $F$, with $n=100,p=300$. As choices of $F$, we consider the multivariate Gaussian (MVG) and $t$-distribution with 3 degrees of freedom $(t_3)$, both having mean $\mu$ and covariance matrix $\Sigma$. We evaluate two choices of $\Sigma$: (i) autoregressive (AR): for $1 \leq i,j \leq p$, we set $(\Sigma)^\text{AR}_{i,j} = 0.5^{|i-j|}$, and (ii) spiked autoregressive (SAR): an otherwise autoregressive $\Sigma$, with the modification $(\Sigma)^\text{SAR}_{i,i} = p+1$ for $i \leq 5$.   Finally, to evaluate the effect of norm choices on the type of alternative hypotheses in calculating our generalized signs, we consider two settings of $\mu$: (a) sparse: $\mu = (\delta/2, \delta, 3\delta/2, 0_{p-3})^T$, and (b) dense: the first 90\% (i.e. 270 of 300) coordinates of $\mu$ is set as the vector $(\delta/2, \delta, 3\delta/2)$, repeated 90 times, and the rest as 0. Depending on $\Sigma$ being AR/SAR, and $\mu$ being sparse/dense, we consider different ranges for $\delta$ for evaluation.

\begin{figure}
\centering
\begin{subfigure}[b]{0.32\textwidth}
\caption{Dense $\mu$, $\Sigma^\text{SAR}$, $a=4,b=4$}
\label{subfig:1a}
\includegraphics[width=\textwidth]{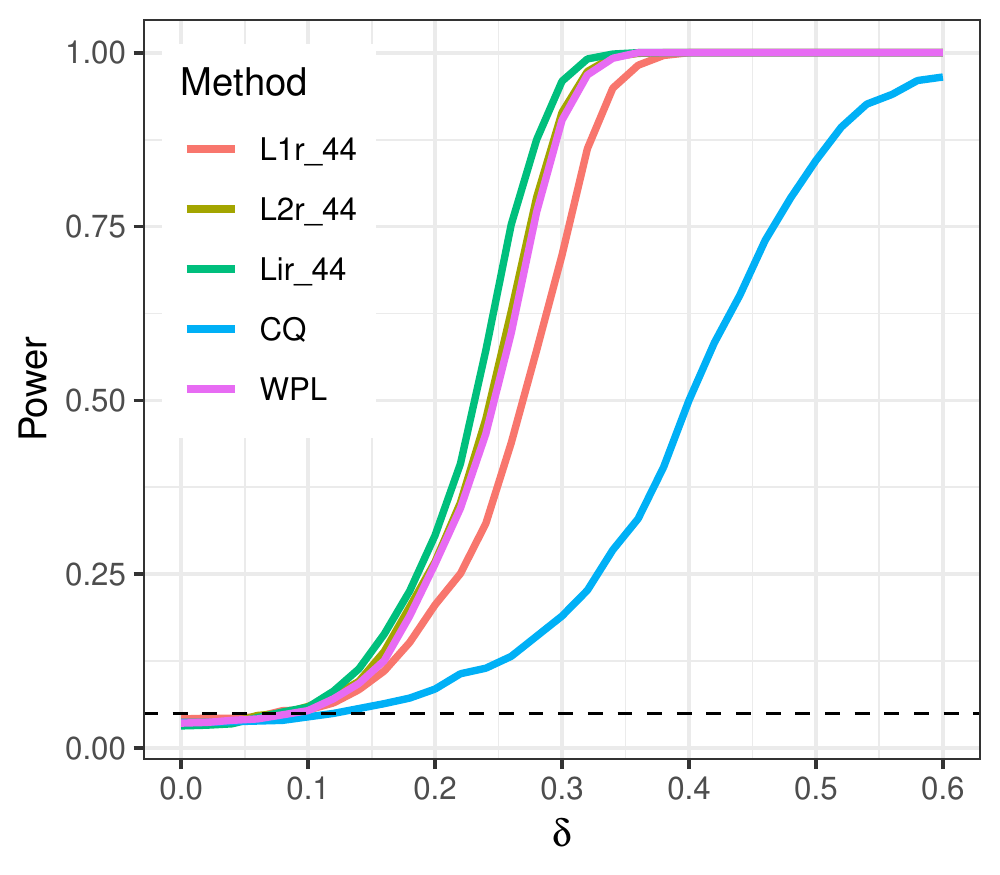}
\end{subfigure}
\begin{subfigure}[b]{0.32\textwidth}
\caption{Sparse $\mu$, $\Sigma^\text{AR}$, $a=4,b=p$}
\label{subfig:1b}
\includegraphics[width=\textwidth]{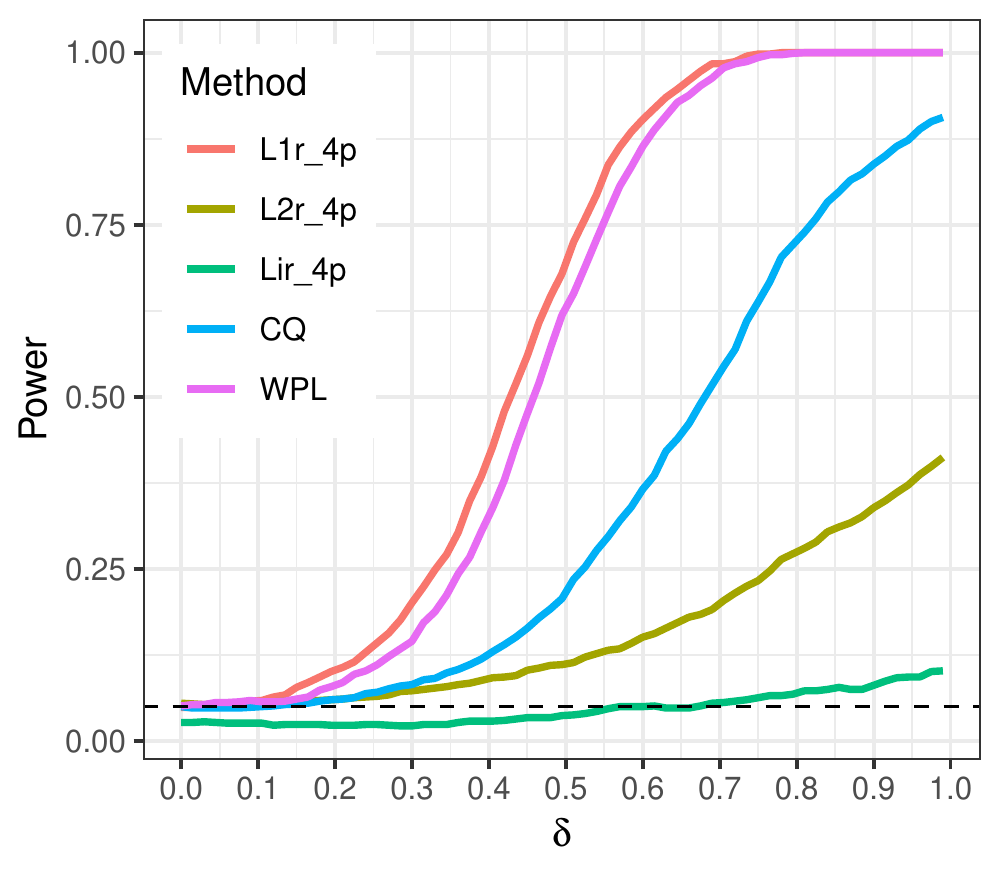}
\end{subfigure}
\begin{subfigure}[b]{0.32\textwidth}
\caption{Dense $\mu$, $\Sigma^\text{SAR}$, $a=1, b=1$}
\label{subfig:2a}
\includegraphics[width=\textwidth]{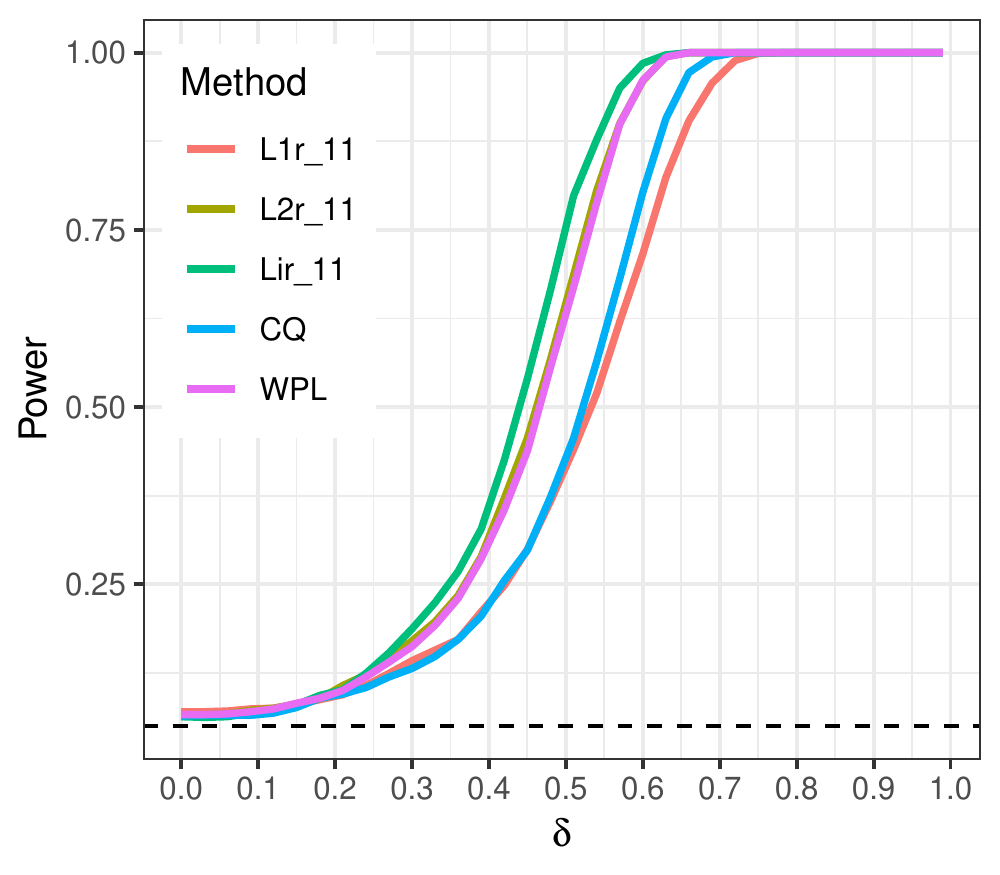}
\end{subfigure}

\caption{Performance comparison on experimental data. Panels (a) and (b) show one-sample test outputs, with a $t_3$ base distribution. Panel (c) shows outputs for two-sample test, with MVG base distribution. Dotted lines indicate nominal size $\alpha=0.05$.}
\label{fig:sims}
\end{figure}

We present the results for two of the above settings in Fig.~\ref{fig:sims} (see Appendix for complete results). The outputs are compared against two existing high-dimensional tests, Chen and Qin \citep[CQ]{ref:AoS10808_HDTest_ChenQin} and Wang et al \citep[WPL]{ref:JASA151658_HDTest_Lan}. We choose these two tests specifically to demonstrate the advantage of our approach compared to asymptotic tests that use similar U-statistics but constructed using inner products of untransformed high-dimensional vectors (CQ test) or $\ell_2$-multivariate signs (WPL test). The difference in performances based on type of alternative evaluated is evident in Fig.~\ref{fig:sims}. When $\mu$ is dense, a generalized sign-based test using the $\ell_\infty$ norm performs the best (Fig.~\ref{subfig:1a}), while for a sparse $\mu$ the $\ell_1$ norm based test works best (Fig.~\ref{subfig:1b}). In general, using the multivariate sign transformation, either with our tests or the WPL test, produces better power performance than the untransformed version (CQ test). Finally, all tests are well-calibrated, i.e. maintain a size close to $\alpha=0.05$ at $\delta=0$.

\paragraph{Two-sample tests}
For this setting, we draw $n_1=60, n_2=40$ samples, respectively from $F_1, F_2$ that have means $\mu_1 = 0_p, \mu_2 \neq 0_p$, and a common covariance matrix $\Sigma$. We consider the same choices of AR or SAR $\Sigma$, sparse or dense $\mu$, and MVG or $t_3$ base distributions as the one-sample case. Results for one of the settings are summarized in Fig.~\ref{subfig:2a}. As we observed in the one-sample case, for a dense $\mu$ the $\ell_\infty$ norm-based generalized sign test has the best performance of all, and the CQ test using the untransformed U-statistic performs the worst.

\begin{table}[t]
\centering
\scalebox{.9}{
\begin{tabular}{cccc|cccc}
\toprule
Dense $\mu$, SAR & $\ell_1$ & $\ell_2$ & $\ell_\infty$ &
Sparse $\mu$, AR & $\ell_1$ & $\ell_2$ & $\ell_\infty$ \\
$\delta$ & ON/RN & ON/RN & ON/RN & $\delta$ & ON/RN & ON/RN & ON/RN \\\midrule
0.00 & 0.06/0.07 & 0.06/0.06 & 0.06/0.06 & 0.00 & 0.05/0.05 & 0.05/0.05 & 0.05/0.05 \\ 
0.15 & 0.06/0.08 & 0.07/0.08 & 0.07/0.08 & 0.30 & 0.09/0.09 & 0.09/0.09 & 0.08/0.08 \\ 
0.30 & 0.11/0.14 & 0.15/0.17 & 0.17/0.19 & 0.60 & 0.25/0.25 & 0.25/0.25 & 0.25/0.24 \\ 
0.45 & 0.23/0.30 & 0.41/0.46 & 0.50/0.54 & 0.90 & 0.63/0.63 & 0.63/0.62 & 0.60/0.61 \\ 
0.60 & 0.54/0.72 & 0.95/0.96 & 0.98/0.98 & 1.20 & 0.94/0.94 & 0.94/0.94 & 0.92/0.92 \\ 
0.75 & 0.97/1.00 & 1.00/1.00 & 1.00/1.00 & 1.50 & 1.00/1.00 & 1.00/1.00 & 0.99/0.99 \\ 
\bottomrule
\end{tabular}
}
\caption{Rejection proportions for two-sample generalized sign tests, using three norm choices, computed using Oracle null (ON) or randomization null (RN). Kernel parameters are fixed at $a=1,b=1$.}
\label{tab:comptable}
\end{table}

\paragraph{Quality of null approximation}
To evaluate the quality of approximating the null hypothesis using our randomization procedure, we generate the null distributions for our one-sample and two-sample test statistics, by setting $\delta=0$, then using Monte-Carlo sampling of size $S = 5000$ (i.e. draw 5000 one-sample or two-sample datasets in each setting and calculate test statistic). Table~\ref{tab:comptable} presents a comparison of two-sample test powers/sizes for different values of $\delta$ and dense or sparse $\mu$, calculated using tail probabilities of this `oracle' null distribution, and compares against the corresponding values obtained by our randomization procedure. As demonstrated previously by \cite{WANG2019160},  quality of Monte Carlo approximation for the null distribution deteriorates in presence of a spiked covariance matrix (left half of Table~\ref{tab:comptable}). Consequently, the ON power performance takes a hit for $\ell_1$ norm, and to some extent the $\ell_2$ and $\ell_\infty$ norms. The RN powers are comparatively robust to this phenomenon, and maintain otherwise similar values to ON across different tests.

\begin{figure}
\centering
\begin{subfigure}[b]{0.32\textwidth}
\caption{Dense $\mu$, $\Sigma^\text{SAR}$, $a=4,b=4$}
\label{subfig:ka}
\includegraphics[width=\textwidth]{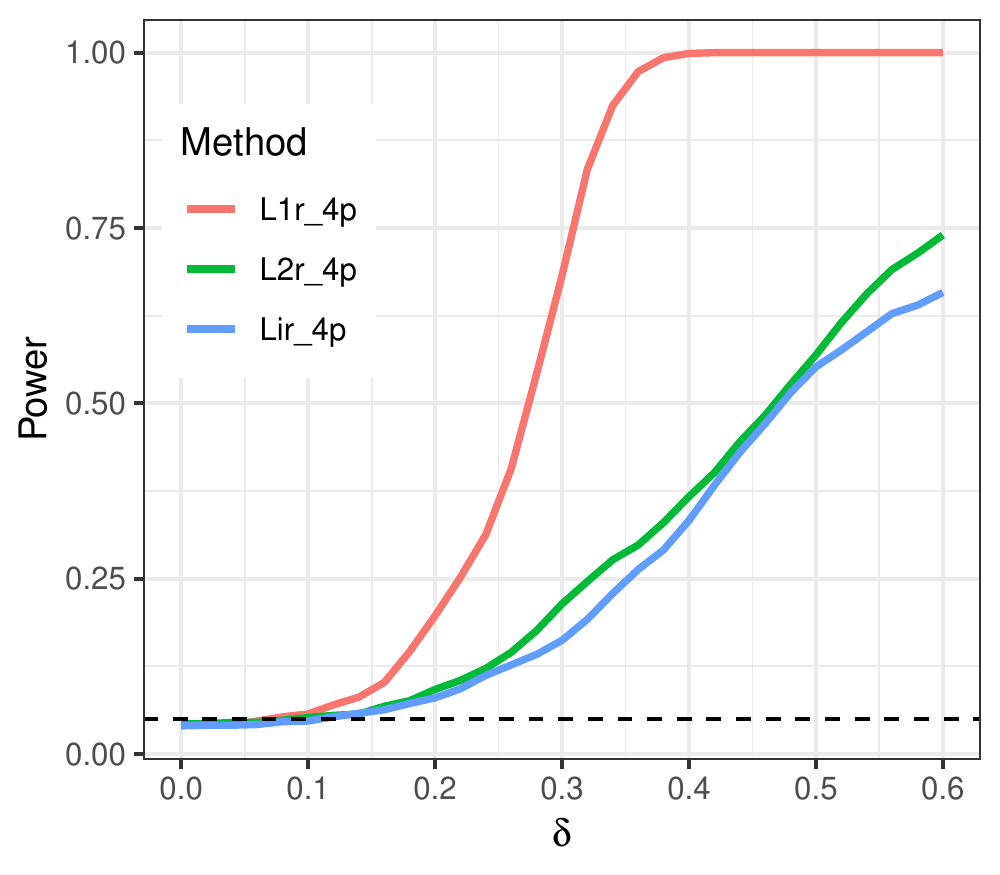}
\end{subfigure}
\begin{subfigure}[b]{0.32\textwidth}
\caption{Sparse $\mu$, $\Sigma^\text{AR}$, $a=4,b=p$}
\label{subfig:kb}
\includegraphics[width=\textwidth]{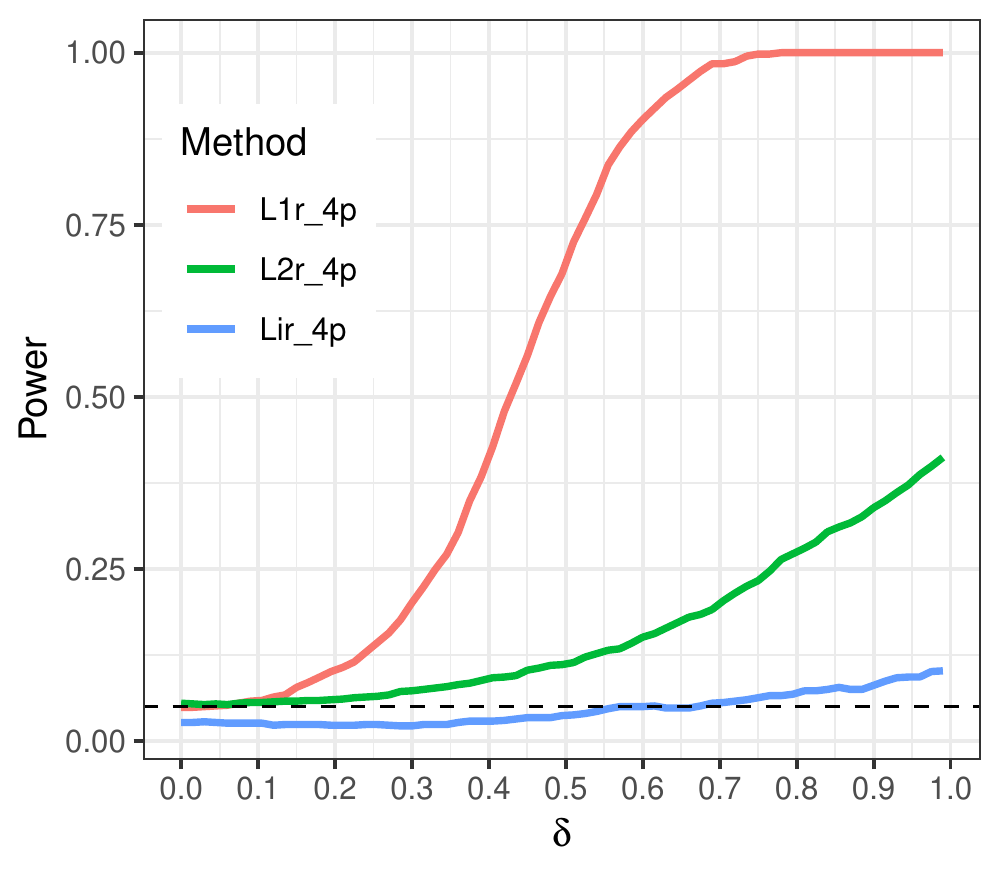}
\end{subfigure}
\begin{subfigure}[b]{0.32\textwidth}
\caption{Sparse $\mu$, $\Sigma^\text{AR}$, $a=p,b=p$}
\label{subfig:kc}
\includegraphics[width=\textwidth]{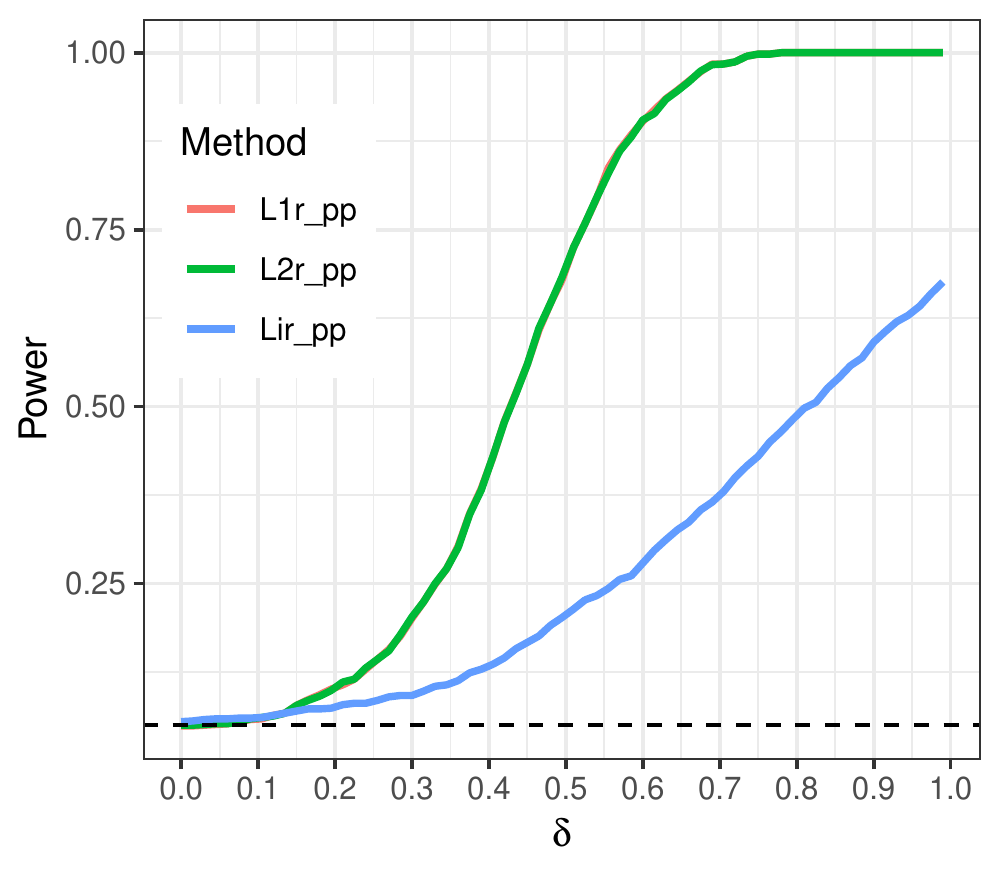}
\end{subfigure}

\caption{Effect of kernel choice on power performance. Dotted lines indicate nominal size $\alpha=0.05$.}
\label{fig:kern}
\end{figure}

\paragraph{Choice of kernel parameters}
We present a number of comparisons for the effect of choosing parameters of the polynomial kernel on test performance in Figure~\ref{fig:kern}. While the performances for other choices of $(a,b)$ we consider, namely $(1,1),(4,4),(p,4)$, remain about the same, combining $b=p$ with a choice of $\ell_2$ or $\ell_\infty$ norm results in deterioration of performance. This is more severe for the sparse $\mu$ case (Figs.~\ref{subfig:kb},~\ref{subfig:kc}).

\section{Real data}

\paragraph{MNIST digits}
As our first real data example, we apply generalized sign based tests to perform pairwise comparison between groups of images of different digits from the publicly available MNIST database \cite{mnist}. To this end, we convert each sample image into a 784 length vector of pixel values. For each pair of image groups, the null hypothesis is that two groups of images are of the same digit, and the alternate hypothesis is that they are not. Intuitively, only a few critical parts in the full image are instrumental behind distinguishing between images of two digits, so it is likely that true alternatives for these pairwise comparisons are sparse. We apply the two-sample generalized sign tests based 3 norm choices, as well as two-sample versions of CQ and WPL tests for this purpose.

\begin{figure}[t]
\centering
\begin{subfigure}[b]{0.3\textwidth}
\caption{$\ell_1$ norm, $a=4, b=p$}
\label{subfig:m1}
\includegraphics[width=\textwidth]{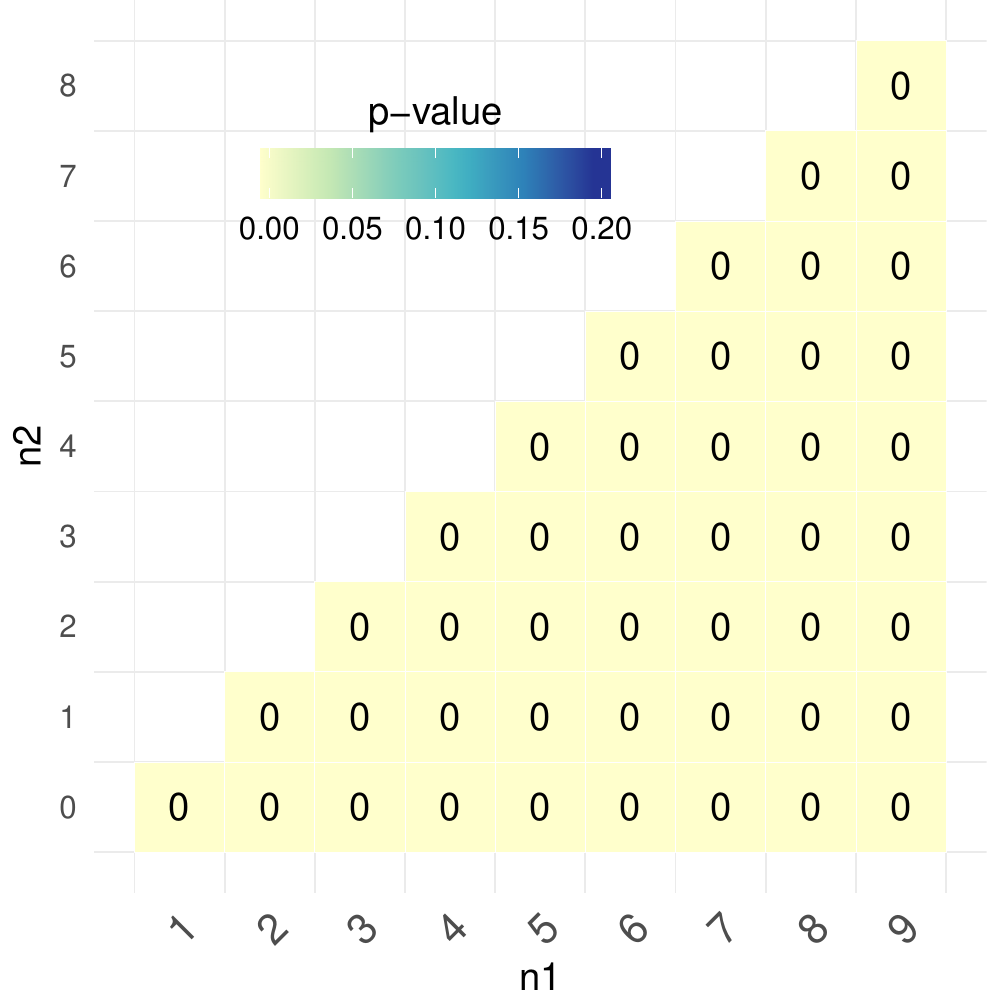}
\end{subfigure}
\begin{subfigure}[b]{0.3\textwidth}
\caption{$\ell_2$ norm, $a=4, b=p$}
\label{subfig:m2}
\includegraphics[width=\textwidth]{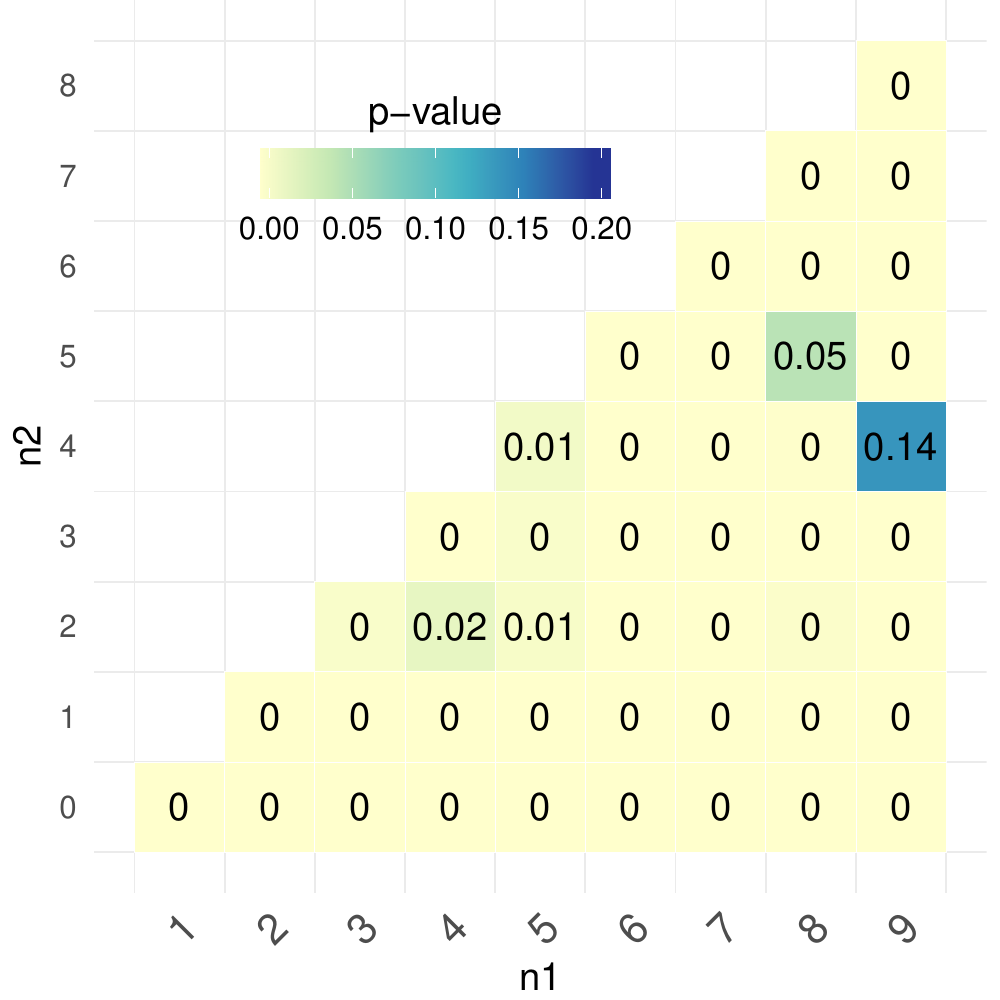}
\end{subfigure}
\begin{subfigure}[b]{0.3\textwidth}
\caption{$\ell_\infty$ norm, $a=4, b=p$}
\label{subfig:m3}
\includegraphics[width=\textwidth]{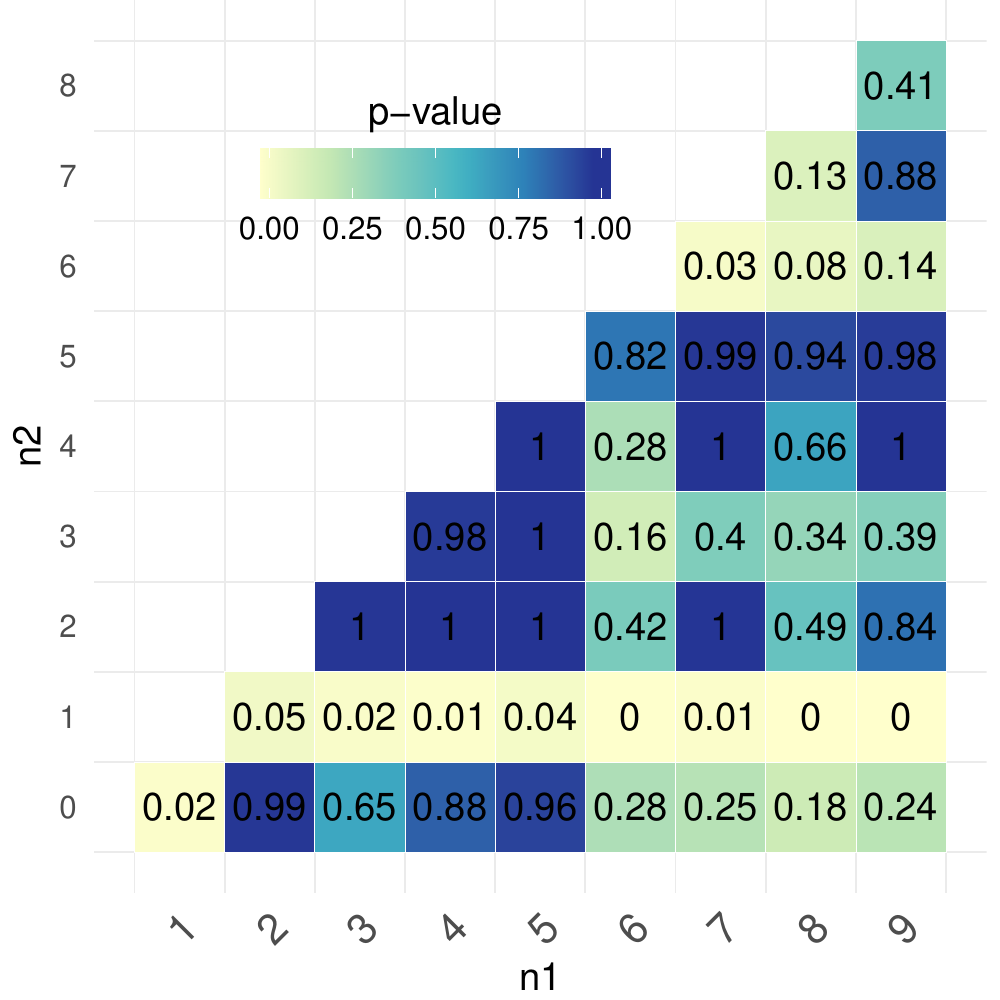}
\end{subfigure}

\caption{Outputs on MNIST dataset: $p$-values from pairwise comparison of digits.}
\label{fig:mnist}
\end{figure}

Figure~\ref{fig:mnist} presents results for three of the applied methods. As expected, the $\ell_\infty$ version performs poorly, and fails to detect differences between many of the digits. The $\ell_2$ version does quite well, except for the 9-vs-4 and 5-vs-8 comparisons. The $p$-value for 5-vs-8 is borderline (0.05), while it fails to distinguish between 9 and 4 ($p$-value 0.14). This is a known issue with the MNIST dataset and has been recognized extensively in previous work. In comparison, the $\ell_1$ version of our test produces very low $p$-values for all comparisons, and thus performs best among all three. In the full results in the Appendix, the WPL test does marginally worse than the $\ell_2$ test, and the non-kernel version of the $\ell_1$ test does marginally better. All polynomial kernel choices for the $\ell_1$ test do better than the linear kernel, and are able to distinguish between each pair of digits.

\paragraph{Twin Studies}
This Genome Wide Association Studies (GWAS) dataset is due to Minnesota Center for Twin and Family Research \cite{MillerEtal12}, and has been researched extensively to determine genetic factors behind quantitative behavioral traits like alcoholism or drug abuse. Detecting effects of SNPs on such traits typically requires large sample sizes, and is difficult due to weak signals of individual SNPs and correlation between their expressions. The curated and anonymized dataset we analyze consists of gene expression data on 527829 SNPs from 7694 individuals, and their recorded severity scores for a number of behavioral traits. We focus on a number of genes related to alcoholism, and consider the problem of detecting the effect of SNPs situated inside these genes on other traits as well. To test for each trait, we compare the expression of these SNPs between individuals in the top and bottom $k$-percentiles of the trait score, with $k = 1, 5, 10$.

Table~\ref{tab:snptable} presents our results, obtained using the $\ell_1$ norm non-kernel version of the two-sample generalized sign test. We identify association between almost all (except OPRM1) genes alcohol consumption/dependence, with significant or borderline $p$-values across one of more quantile comparison. In addition, we detect a number of associations with traits not related to alcoholism. For example, the gene DRD2 exhibits significant different expressions when top and bottom 5\% quantiles are compared. This is corroborated by the well-documented effect of dopamine receptor activation on behavioral disinhibition \cite{drd2ref}, and the fact that DRD2 encodes the D2 subtype of the dopamine receptor protein in our body. In addition to alcoholism, mirroring our findings previous studies have documented the association of the expression of GABRA2 with drug abuse \cite{EnochEtal11}, behavioral disinhibition \cite{WangEtal16}, and externalizing \cite{DickEtal09}.

\begin{table}[t]
\centering
\scalebox{.9}{
\begin{tabular}{rcccc}
\toprule
Trait & GABRA2 & ADH1--7 & SLC6A3 & SLC6A4 \\
\midrule
Nicotine addiction  & 0.2/0.95/0.53 & 0.19/0.53/0.88 & 0.43/0.12/0.19 & 0.77/0.18/0.92 \\
Alcohol consumption & {\bf 0.05}/0.7/{\bf 0.04} & {\it 0.07}/0.52/0.25 & 0.59/0.33/{\bf 0.04} & 0.36/0.7/{\it 0.09} \\
Alcohol dependence & 0.08/0.58/0.09 & {\bf 0.03}/0.26/0.24 & 0.31/0.13/0.71 & 0.61/0.27/{\it 0.1} \\
Drug abuse & {\bf 0.05}/0.63/0.28 & 0.07/0.18/0.5 & 0.21/0.69/0.7 & {\bf 0.03}/0.43/0.7 \\
Behavioral Disinhibition & 0.7/0.13/{\it 0.09} & 0.71/0.19/0.5 & 0.17/0.44/0.73 & 0.12/0.84/0.71 \\
Externalizing & 0.73/{\bf 0.02}/0.26 & {\it 0.1}/0.23/0.31 & 0.22/0.83/0.58 & 0.62/0.43/0.85 \\
\midrule
Trait & OPRM1 & DRD2 & ALDH2 &\\
\midrule
Nicotine addiction & 0.97/0.88/0.76 & 0.29/0.49/0.91 & 0.76/0.54/0.9 &\\
Alcohol consumption & 0.3/0.16/0.14 & 0.66/{\it 0.09}/0.6 & 0.43/{\it 0.07}/{\it 0.07} &\\
Alcohol dependence & 0.43/0.56/0.94 & 0.12/0.88/0.84 & 0.15/0.78/0.89 &\\
Drug abuse & 0.45/0.24/0.24 & 0.52/0.62/0.69 & 0.14/{\bf 0.02}/0.49 &\\
Behavioral Disinhibition & 0.42/0.58/0.4 & 0.81/{\bf 0.03}/{\it 0.09} & 0.59/0.91/0.85 &\\
Externalizing & 0.47/0.49/0.48 & {\it 0.06}/0.2/0.5 & 0.13/0.62/{\it 0.06} &\\
\bottomrule
\end{tabular}
}
\caption{Trait and gene-specific $p$-values from the Twin Studies data, obtained by comparing individuals of top and bottom 1, 5, and 10\% quantiles for each trait score. Significant ($\leq 0.05$) and borderline ($\geq 0.05, \leq 0.1$) $p$-values are indicated by bold and italics, respectively.}
\label{tab:snptable}
\end{table}

\section{Conclusion}
In this paper, we present a general framework for nonparametric high-dimensional hypothesis tests in one- and two-sample setups. Based on our proposed generalized multivariate sign transformation, we develop a broad set of theoretical tools that derive dimensional asymptotics on performance guarantees of the resulting tests, using minimal assumptions on the underlying data distributions. By tuning a number of norm and kernel choices, it is possible to incorporate information on the geometry of the parameter space, thereby tailoring the tests to maximize power for specific alternate hypotheses. Future work in this direction include a principled way of calibrating the choice of kernel parameters, as well as generalizing to manifold or graph-valued data using underlying properties of the manifold geometry or graph laplacian (e.g. \cite{rustamov2021intrinsic}) to adapt our theoretical tools to such settings.

\bibliographystyle{plain}
\bibliography{gsign}

\begin{thebibliography}{10}

\bibitem{ref:Anderson_Multivariate_Book}
T.~W. Anderson.
\newblock {\em An Introduction To Multivariate Statistical Analysis}.
\newblock Wiley New York, 3 edition, 2003.

\bibitem{ref:StatSinica96311_HDTest_BaiSaranadasa}
Z.~Bai and H.~Saranadasa.
\newblock Effect of high dimension: By an example of a two sample problem.
\newblock {\em Statistica Sinica}, pages 311--329, 1996.

\bibitem{ref:JRSSB14349_HDTest_Cai}
T.~T. Cai, W.~Liu, and Y.~Xia.
\newblock Two-sample test of high dimensional means under dependence.
\newblock {\em Journal of the Royal Statistical Society: Series B: Statistical
  Methodology}, pages 349--372, 2014.

\bibitem{ref:AoS17771_HDTest_Probal}
A.~Chakraborty, P.~Chaudhuri, et~al.
\newblock Tests for high-dimensional data based on means, spatial signs and
  spatial ranks.
\newblock {\em The Annals of Statistics}, 45(2):771--799, 2017.

\bibitem{ref:AoS10808_HDTest_ChenQin}
S.-X. Chen, Y.-L. Qin, et~al.
\newblock A two-sample test for high-dimensional data with applications to
  gene-set testing.
\newblock {\em The Annals of Statistics}, 38(2):808--835, 2010.

\bibitem{DickEtal09}
D.~M. Dick et~al.
\newblock {Role of GABRA2 in trajectories of externalizing behavior across
  development and evidence of moderation by parental monitoring}.
\newblock {\em Arch. Gen. Psychiatry}, 66(6):649--657, 2009.

\bibitem{EnochEtal11}
M.-A. Enoch et~al.
\newblock {The Influence of GABRA2, Childhood Trauma and their Interaction on
  Alcohol, Heroin and Cocaine Dependence}.
\newblock {\em Biol. Psychiatry}, 67(1):20--27, 2010.

\bibitem{ref:JASA16721_HDTest_Changliang}
L.~Feng, C.~Zou, and Z.~Wang.
\newblock Multivariate-sign-based high-dimensional tests for the two-sample
  location problem.
\newblock {\em J. Amer. Statist. Assoc.}, 111(514):721--735, 2016.

\bibitem{mmd}
A.~Gretton, K.~M. Borgwardt, M.~J. Rasch, B.~Sch\"{o}lkopf, and A.~Smola.
\newblock A kernel two-sample test.
\newblock {\em J. Mach. Learn. Res.}, 13:723--773, March 2012.

\bibitem{mnist}
Y.~LeCun, C.~Cortes, and C.~Burges.
\newblock {The MNIST database of handwritten digits}.

\bibitem{ref:AoS99783_Depth_LiuSingh}
R.~Y. Liu, J.~M. Parelius, and K.~Singh.
\newblock Multivariate analysis by data depth: Descriptive statistics, graphics
  and inference, (with discussion).
\newblock {\em The Annals of Statistics}, 27(3):783--858, 1999.

\bibitem{ManolioEtal09}
T.~A. Manolio, F.~S. Collins, N.~J. Cox, et~al.
\newblock {Finding the missing heritability of complex diseases}.
\newblock {\em Nature}, 461:747--753, 2009.

\bibitem{MillerEtal12}
M.~B. Miller, S.~Basu, J.~Cunningham, et~al.
\newblock {The Minnesota Center for Twin and Family Research Genome-Wide
  Association Study}.
\newblock {\em Twin Res Hum Genet.}, 15:767--774, 2012.

\bibitem{ref:Muirhead_Multivariate_Book}
R.~J. Muirhead.
\newblock {\em Aspects Of Multivariate Statistical Theory}, volume 197.
\newblock John Wiley \& Sons, 2009.

\bibitem{rustamov2021intrinsic}
R.~M. Rustamov and S.~Majumdar.
\newblock Intrinsic sliced wasserstein distances for comparing collections of
  probability distributions on manifolds and graphs.
\newblock {\em arXiv: 2010.15285}, 2021.

\bibitem{ref:DIMACS061_Serfling}
R.~Serfling.
\newblock Depth functions in nonparametric multivariate inference.
\newblock {\em DIMACS Series in Discrete Mathematics and Theoretical Computer
  Science}, 72:1, 2006.

\bibitem{serflingbook}
R.~J. Serfling.
\newblock {\em {Approximation theorems of mathematical statistics}}.
\newblock {John Wiley \& Sons}, 2009.

\bibitem{SpatialSign}
S.~Serneels, E.~{de Nolf}, and P.~J. {Van Espen}.
\newblock {Spatial sign preprocessing: a simple way to impart moderate
  robustness to multivariate estimators}.
\newblock {\em J Chem Inf Model}, 46(3):1402--1409, 2006.

\bibitem{ref:JMVA09518_HDTest_Srivastava}
M.~S. Srivastava.
\newblock A test for the mean vector with fewer observations than the dimension
  under non-normality.
\newblock {\em Journal of Multivariate Analysis}, 100(3):518--532, 2009.

\bibitem{ref:JMVA08386_HDTest_SrivastavaDu}
M.~S. Srivastava and M.~Du.
\newblock A test for the mean vector with fewer observations than the
  dimension.
\newblock {\em Journal of Multivariate Analysis}, 99(3):386--402, 2008.

\bibitem{drd2ref}
M.M. van Gaalen et~al.
\newblock {Behavioral disinhibition requires dopamine receptor activation}.
\newblock {\em Psychopharmacology (Berl)}, 187(1):73--85, 2006.

\bibitem{WangEtal16}
F.~L. Wang et~al.
\newblock {Mechanisms in the relation between GABRA2 and adolescent
  externalizing problems}.
\newblock {\em Eur. Child Adolesc. Psychiatry}, 25(1):67--80, 2016.

\bibitem{ref:JASA151658_HDTest_Lan}
L.~Wang, B.~Peng, and R.~Li.
\newblock A high-dimensional nonparametric multivariate test for mean vector.
\newblock {\em Journal of the American Statistical Association},
  110(512):1658--1669, 2015.

\bibitem{WANG2019160}
R.~Wang and X.~Xu.
\newblock A feasible high dimensional randomization test for the mean vector.
\newblock {\em Journal of Statistical Planning and Inference}, 199:160--178,
  2019.

\bibitem{ref:AoS00461_ZuoSerfling}
Y.~Zuo and R.~Serfling.
\newblock General notions of statistical depth function.
\newblock {\em The Annals of Statistics}, 28(2):461--482, 2000.

\end{thebibliography}

\section*{Appendix}
\setcounter{section}{0}
\renewcommand\thesection{\Alph{section}}
\section{Proofs of theoretical results}
\label{sec:proofs}

\begin{proof}[Proof of Theorem~1]

For convenience, let us use the notations $\Delta = \BE D (X_{1})$ and 
$\gamma = \bigl(\mu^{T} \Sigma \mu \bigr)^{1/2}$.  
We assume that $p = o ( \Delta^{4})$, this is trivially satisfied when $D (\cdot)$ is $\ell_{k}$-norm for $k = 1, 2, \infty$. We also assume that
$\epsilon_{n} = n^{-8/p}$. 

Recall that 
\baq 
T_{D, n} = {\binom{n}{2}}^{-1} \sum_{1 \leq i < j \leq n} S_{i}^{T} S_{j}. 
\label{eq:T_E}
\eaq
where 
\baq 
S_{i} = {\frac {X_{i}}{D (X_{i})}} \cI_{\{ D (X_{i}) > \epsilon_{n} \}}.
\label{eq:S} 
\eaq

We recall our technical conditions here for convenience of presentation.
The $X_{i}$'s are  iid copies of some absolutely continuous random variable $X \in \cX \subseteq \BR^{p}$ such that
\baq 
\BE X_{1} & = \mu \in \BR^{p},  \label{eq:mu}\\
\BV X_{1} & = \Sigma > 0,  \label{eq:Sigma}\\
\BE | X_{1} |^{8} & = O (p) < \infty.  \label{eq:XMoment}
\eaq
We assume that that the eigenvalues of $\Sigma$ are bounded above and below
\baq 
\lambda_{j} (\Sigma) \in (C^{-1}, C) \ \ \text{ for some } \ C > 0, 
\text{ for all } j = 1, \ldots, p. 
\label{eq:Eigenvalue}
\eaq
Also, we will use the 
notation $C$, with or without subscripts, as generic for constants below, and these do not depend on $n$ or $p$.
We assume that for sufficiently small
$\epsilon > 0$ we have 
\baq
\bigl\{x: D (x) \leq \epsilon \bigr\}
\subseteq 
\bigl\{x: |x|_{1} \leq C_{0} \epsilon \bigr\}, 
\label{eq:DCondition}
\eaq
for some $C_{0} > 0$. We assume that the density $f (\cdot)$ of $X$ satisfies the following condition:
There exists a $\delta > 0$ and a positive constant $C_{1} > 0$ such that
\baq
f (x) \leq C_{1}, \ \ \text{ for all } |x|_{1} < \delta, \text{ and }
\label{eq:Density} \\
\BE \bigl( D (X_{i}) - \BE D (X_{i}) \bigr)^{8} | X_{i}|^{8} = O (p). 
\label{eq:DZ1}
\eaq


Define the event 
\ban 
\cA_{i} = \bigl\{  D (X_{i})  > \epsilon_{n} \bigr\}. 
\ean 
Our first task is to show that $\cA_{i}$ is a ``large'' set. Note that since 
$\{ \epsilon_{n} \}$ is a decreasing sequence, for sufficiently large $n$ we have 
\ban 
\cA_{i}^{C} & = \bigl\{  D (X_{i})  \leq \epsilon_{n} \bigr\} \\
& \subseteq 
\bigl\{x: |x|_{1} \leq C_{0} \epsilon_{n} \bigr\}
\ean 
using \eqref{eq:DCondition}. Hence, using \eqref{eq:Density}, we have 
\baq 
\BP \bigl[  \cA_{i}^{C}  \bigr]
= O ( \epsilon_{n}^{p} ). 
\label{eq:AComplementProbability}
\eaq

Define 
\ban 
R_{i} = {\frac {D (X_{i}) - \BE D (X_{i})}{\BE D (X_{i}) }}, \ i = 1, \ldots, n.
\ean 
Then
\ban 
S_{i} & = {\frac {X_{i}}{D (X_{i})}} \cI_{\{ D (X_{i}) > \epsilon_{n} \}} \\
 & = {\frac {X_{i}}{ \BE D (X_{i})  (1 + R_{i})}} \cI_{\{ D (X_{i}) > \epsilon_{n} \}} \\
& = {\frac {X_{i}}{\BE D (X_{i})}} \Bigl( 1 - R_{i} + {\frac{R_{i}^{2}}{1 + R_{i}}} \Bigr)
  \cI_{\{ D (X_{i}) > \epsilon_{n} \}} \\ 
& = \Delta^{-1} X_{i} \Bigl( 1 - R_{i} + {\frac{R_{i}^{2}}{1 + R_{i}}} \Bigr)
  \cI_{\{ D (X_{i}) > \epsilon_{n} \}} \\ 
 \ean
 
 Consequently we have for $i \ne j$
 \ban 
 S_{i}^{T} S_{j} 
 & = \Delta^{-2}  (X_{i}^{T} X_{j})
 \Bigl( 1 - R_{i} + {\frac{R_{i}^{2}}{1 + R_{i}}} \Bigr) 
 \Bigl( 1 - R_{j} + {\frac{R_{j}^{2}}{1 + R_{j}}} \Bigr)
   \cI_{\{\cA_{i} \cap \cA_{j} \}} \\
 & =   T_{ij} + \sum_{a = 1}^{9} R_{a i j}, \ \text{ where}  \\
  T_{ij} & = \Delta^{-2} \  (X_{i}^{T} X_{j}), \\
  R_{1 i j} & = \Delta^{-2}\   (X_{i}^{T} X_{j}) \ 
  \cI_{\{\cA_{i}^{C} \cup \cA_{j}^{C} \}}, \\
  R_{2 i j} & = -  \Delta^{-2} \ R_{i} (X_{i}^{T} X_{j}) \ 
   \cI_{\{\cA_{i} \cap \cA_{j} \}} \\
  R_{3 i j} & = - \Delta^{-2} \ R_{j}   (X_{i}^{T} X_{j}) \ 
   \cI_{\{\cA_{i} \cap \cA_{j} \}} \\
  & \ \text{ and so on.}
  \ean
  
  We now establish the properties of each of these ten quantities. 
Thus we have 
\ban 
T_{ij} & = \Delta^{-2} X_{i}^{T} X_{j} \\
& = \Delta^{-2}  \Bigl(  | \mu|^{2} + \mu^{T} \Sigma^{1/2} (Z_{i} + Z_{j})
+ Z_{i}^{T} \Sigma Z_{j} \Bigr) \\
& = T_{1 ij} + T_{2 ij} + T_{3ij}. 
\ean 
Here, $T_{1 ij} \equiv T_{1} = \Delta^{-2} | \mu|^{2}$ which does not depend on 
$i, j$ (and hence remains as it is in $T$ in \eqref{eq:T_E}), and is $T_{1} = 0$ under 
the null hypothesis $\mu = 0$ but is the leading non-centrality term under the alternative 
$\mu \ne 0$. 

We now show properties of the rest of the terms. Note that $T_{2 ij}$ and $T_{3 ij}$ both have mean zero. Note that
 \ban 
T_{2} & =  {\binom{n}{2}}^{-1} \sum_{1 \leq i < j \leq n} 
  T_{2 ij}
   = 2 \Delta^{-2}  n^{-1} \sum_{i = 1}^{n} 
\mu^{T} \Sigma^{1/2} Z_{i}. 
\ean 
Hence 
\ban 
\BV T_{2} = 4 \Delta^{-4}  n^{-1} \gamma^{2}. 
\ean
Similarly,
\ban
T_{3} & =  \Delta^{-2} {\binom{n}{2}}^{-1} \sum_{1 \leq i < j \leq n}   T_{3 ij} \\
& =  \Delta^{-2}  {\binom{n}{2}}^{-1} \sum_{1 \leq i < j \leq n} Z_{i}^{T} \Sigma Z_{j}
\ean
 and since $ \BV T_{3 ij}  = \Delta^{-4} \trace (\Sigma^{2})$, we have that
\ban 
 \BV T_{3} & = C_{2} n^{-2}\Delta^{-4} \trace (\Sigma^{2}). 
 \ean
 We can now see that 
 \ban 
T_{2} =  {\binom{n}{2}}^{-1} \sum_{1 \leq i < j \leq n} 
  T_{2 ij} =  O_{P} (n^{-1/2} \Delta^{-2}  \gamma ), \\
T_{3} =  {\binom{n}{2}}^{-1} \sum_{1 \leq i < j \leq n} 
  T_{3 ij} =  O_{P} (n^{-1} \Delta^{-2} \trace (\Sigma^{2}) ). 
  \ean
  Notice that under condition \eqref{eq:Eigenvalue}, 
  we have $\trace (\Sigma^{2}) ) = O(p)$. Also, 
  if $D (x) = |x|_{1}$, then $\Delta = O(p)$, and if $D (x) = |x|_{2}$ or 
  $D (x) = |x|_{\infty}$, $\Delta = O(p^{1/2})$. This means that 
  $T_{3} = O (n^{-1} p^{-1})$ if we use $D (x) = |x|_{1}$, and 
  $T_{3} = O (n^{-1})$ if we use $D (x)$ to be $\ell_{2}$ or $\ell_{\infty}$ norm.   Note that $T_{2} = 0$ under the null, so $T_{3}$ is the main contributant to the rate of convergence of $T$ (under null), and this can be $np$ if $D (x) = |x|_{1}$. Under $D (x) = |x|_{2}$ or   $D (x) = |x|_{\infty}$, we get the well-known rate of $n$ for degenerate $U$-statistics.

Let us now look at 
\ban 
\BE 
 R_{1 i j}^{2} & = \BE \biggl[ 
  \Delta^{-2} X_{i}^{T} X_{j}
  \cI_{\cA_{i}^{C} \cup \cA_{j}^{C}} \biggr]^{2} \\
  & \leq K \Delta^{-4} \BP^{1/2} \bigl[  \cA_{1}^{C}  \bigr]
  \Bigl( \BE (X_{i}^{T} X_{j})^{4}  \Bigr)^{1/2} \\
  & =  O ( \Delta^{-4} p \epsilon_{n}^{p/2} ),  
  \ean
 using \eqref{eq:AComplementProbability} and \eqref{eq:XMoment}. 
  
 Consequently, we have 
 \ban  
 R_{1} =  {\binom{n}{2}}^{-1} \sum_{1 \leq i < j \leq n}
  R_{1 i j} = O_{P} ( \Delta^{-2} p^{1/2} \epsilon_{n}^{p/4}).
  \ean
  
 Let us now consider the next term: 
\ban   
  R_{2 i j} & = - R_{i} \Delta^{-2} X_{i}^{T} X_{j}
   \cI_{\{\cA_{i} \cap \cA_{j} \}}  \\
  & =  - R_{i} \Delta^{-2} X_{i}^{T} X_{j}
   + R_{i} \Delta^{-2} X_{i}^{T} X_{j}
   \cI_{\cA_{i}^{C} \cup \cA_{j}^{C}} \\
& = R_{2 1 ij} + R_{22 ij}. 
\ean

Let us look at $R_{2 1 i j}$ carefully now. 
\ban 
R_{2 1 i j} & = - \Delta^{-3} \bigl( D (X_{i}) - \BE D (X_{i}) \bigr) 
\bigl( \mu^{T} + Z_{i}^{T} \Sigma^{1/2} \bigr)
\bigl( \mu + \Sigma^{1/2} Z_{j} \bigr) \\
& = - \Delta^{-3} 
\biggl[ 
\bigl( D (X_{i}) - \BE D (X_{i}) \bigr)
\Bigl\{ |\mu|^{2} + \mu^{T} \Sigma^{1/2} Z_{j}
\Bigr\} 
\\
& \hspace{1cm}
+ \bigl( D (X_{i}) - \BE D (X_{i}) \bigr)
\Bigl\{ \mu^{T} \Sigma^{1/2} Z_{i} + 
Z_{j}^{T} \Sigma Z_{i}
\Bigr\}
\biggr]. 
\ean

Note that if $\mu = 0$, $\BE R_{2 1 i j} = 0$. But when $\mu \ne 0$ , we have 
\ban 
\BE R_{2 1 i j} 
= - \Delta^{-3} \BE \bigl( D (X_{i}) - \BE D (X_{i}) \bigr) \mu^{T} \Sigma^{1/2} Z_{i}. 
\ean
We have 
\ban 
\bigl( \BE R_{2 1 i j} \bigr)^{2}
& \leq \Delta^{-6} \gamma^{2} 
\Bigl( \BE \bigl( D (X_{i}) - \BE D (X_{i}) \bigr)^{2} |Z_{i}|^{2}  \Bigr)^{2} \\
& = O (\Delta^{-6} \gamma^{2} p ),
\ean
using \eqref{eq:DZ1}.

This implies that $R_{2 1 i j}$ contributes a non-centrality term 
$O (\Delta^{-3} \gamma p^{1/2})$ under the alternative. This term is 
$o ( T_{1})$ under the assumption $p = o( \Delta^{4})$, hence this is 
a negligible term. A simple computation shows that 
$\BE R_{2 1 i j}^{2} = O (\Delta^{-6} p )$ generally.


We also have that 
\ban 
\BE R_{22 ij}^{2} & = 
\BE \Bigl(  \Delta^{-3} \bigl( D (X_{i}) - \BE D (X_{i}) \bigr) X_{i}^{T} X_{j}
   \cI_{\cA_{i}^{C} \cup \cA_{j}^{C}} \Bigr)^{2} \\
& \leq K \Delta^{-6}  \BP^{1/2} \bigl[  \cA_{1}^{C}  \bigr]
\biggl( \BE \Bigl(  \bigl( D (X_{i}) - \BE D (X_{i}) \bigr)  X_{i}^{T} X_{j}\Bigr)^{4} \biggr)^{1/2}
& = O \bigl(  \Delta^{-6} p \epsilon_{n}^{p/2} \bigr),  
  \ean
 using \eqref{eq:AComplementProbability} and \eqref{eq:DZ1}.
Thus we have
\ban 
R_{2} =  {\binom{n}{2}}^{-1} \sum_{1 \leq i < j \leq n}
  R_{2 i j}
= O \bigl( \Delta^{-3} \gamma p^{1/2} + \Delta^{-3} p^{1/2} \epsilon_{n}^{p/4}
\bigr). 
\ean

The analysis for $R_{3}$ is identical to that of $R_{2}$. The other terms can also be shown to be negligible, we omit the routine algebraic details.

Consequently, we get the following:
\begin{itemize}[leftmargin=*]
\item The leading random term is 
\ban
T_{3} 
& =  \Delta^{-2}  {\binom{n}{2}}^{-1} \sum_{1 \leq i < j \leq n} Z_{i}^{T} \Sigma Z_{j} \\
& = O_{P} (n^{-1} p \Delta^{-2} ). 
\ean
This can be $np$ if $D (x) = |x|_{1}$, since in that case $\Delta = O(p)$. 
If $D (x) = |x|_{2}$ or  $D (x) = |x|_{\infty}$, $\Delta = O(p^{1/2})$, hence
under these cases we get the well-known rate of $n$ for degenerate $U$-statistics. 
We use 
\ban 
U = n \Delta^{2} p^{-1} T_{E, n}
\ean
as the properly scaled test statistic, and this is $O_{P} (1)$ under the null.

\item The leading non-centrality term is derived from $T_{1} = \Delta^{-2} | \mu|^{2}$,
so this is 
\ban 
\nu = n p^{-1} | \mu|^{2}. 
\ean

\item To make $R_{1}$ (and smaller order terms) negligible, we need 
$ O \bigl( n \Delta^{2} p^{-1}   \Delta^{-2} p^{1/2} \epsilon_{n}^{p/4} \bigr) 
= O \bigl( n p^{-1/2} \epsilon_{n}^{p/4} \bigr) = o (1)$. 
Our choice of $\epsilon_{n} = n^{-8/p}$ ensures that this is the case.

\end{itemize}

In order to construct the limiting distribution, 
for a probability distribution $\BF$ supported on $\BR^{p}$, define the Hilbert spaces
\baq 
L_2 ( \BF ) & = \bigl\{ h: \mathbb{R}^{p} \rightarrow \mathbb{R}, 
	\ \int h^2( x ) d \BF ( x ) < \infty \bigr\}  \text{ and } \\
L_2 ( \BF \times \BF) & = \bigl\{ h: \mathbb{R}^{p} \times \mathbb{R}^{p} \rightarrow \mathbb{R}, 
	\ \int\int h^2( x, y ) d \BF ( x ) d \BF ( y ) < \infty \bigr\}.
 \eaq
 Fix $h \in L_2 ( \BF  \times \BF )$. 
 Define the operator $A_h: L_2 ( \BF ) \rightarrow L_2 ( \BF )$ as
\baq \label{eq:operatordef}
A_{h} g(x) =\int h (x,y) g(y) d \BF ( y ), \ g\in L_2 ( \BF ).
\eaq 
Then
there exists eigenvalues $\{\lambda_j\}$ and corresponding eigenfunctions $\{ \phi_j\} \subset L_{2} ( \BF )$ for the operator $A$. 
That is: 
\baq 
A_{h} \phi_{j} & = \lambda_{j} \phi_{j}, \forall j,  \\
\int \phi_{j}^{2} ( x ) d \BF ( x )  = 1, \ & 
\int \phi_{j} ( x ) \phi_{k} ( x ) d \BF ( x ) = 0,  \ \forall j \ne k, 
\text{ and } 
\label{eq:orthogonal} \\
h ( x, y ) & = \sum_{j = 1}^{\infty} \lambda_{j} \phi_{j} ( x ) \phi_{j}  ( y ). 
\label{eq:spectral}
\eaq 
The equality in \eqref{eq:spectral} is in the $L_2$ sense.
That is, if $Y_1, Y_2$ are i.i.d. $F$ then 
\baq 
\BE [ h ( Y_1, Y_2) - \sum_{k = 1}^n 
\lambda_k \phi_k (Y_1)  \phi_k (Y_2)]^2
\rightarrow 0 \ \ \text{as} \ \ n \rightarrow \infty.
\label{eq:3.47}
\eaq 
Further
\baq
h_1(x) =  \BE h (x,  Y_2) = \sum_{k = 1}^{\infty} \lambda_{k} \phi_{k} ( x ) 
\BE \bigl( \phi_{k} ( Y_2 ) \bigr) \ \ {\text{ almost surely } } \ \ \BF.
\label{eq:h1spectral}
\eaq
Also note that from \eqref{eq:orthogonal} and \eqref{eq:3.47}, 
\baq 
\BE h^2 ( Y_1, Y_2 ) = 
\sum_{ k = 1}^\infty \lambda_k^2.
\eaq 

The rest of the details of the proof follows using the standard non-central limit theorem for degenerate $U$-statistics, Slutsky's Theorem and related results. 
\end{proof}

\begin{proof}[Proof of Theorem 2]
Some of the technical details for this section are similar to those presented in the proof of Theorem 1, consequently we provide the outline of the main arguments here. 

Using a Taylor series expansion, we have 
\ban 
K (\mu + \delta, \nu + \epsilon)
& = K (\mu, \nu + \epsilon) +  K_{1} (\mu, \nu + \epsilon)\delta
+ {\frac{1}{2}} \delta^{T} K_{2} (\mu, \nu + \epsilon) \delta 
+ R_{1} (\delta, \epsilon,  \mu, \nu) \\
& = K (\nu + \epsilon, \mu) +  K_{1} (\mu, \nu + \epsilon)\delta
+ {\frac{1}{2}} \delta^{T} K_{2} (\mu, \nu + \epsilon) \delta 
+ R_{1} (\delta, \epsilon, \mu, \nu) \\
& = K (\nu, \mu) +  K_{1}^{T} (\mu, \nu + \epsilon)\delta
+ {\frac{1}{2}} \delta^{T} K_{2} (\mu, \nu + \epsilon) \delta 
+ R_{1} (\delta, \epsilon,  \mu, \nu) 
\\ & \hspace{1cm}
+ K_{1}^{T} (\nu, \mu) \epsilon
+ {\frac{1}{2}} \epsilon^{T} K_{2} (\nu, \mu) \epsilon 
+ R_{2} (\delta, \epsilon, \mu, \nu) \\ 
& = K (\nu, \mu) + K_{1}^{T} (\mu, \nu)\delta
+ \epsilon^{T}  K_{11} (\mu, \nu)\delta
\\ & \hspace{1cm}
+ {\frac{1}{2}} \delta^{T} K_{2} (\mu, \nu) \delta  
+ K_{1}^{T} (\nu, \mu) \epsilon
+ {\frac{1}{2}} \epsilon^{T} K_{2} (\nu, \mu) \epsilon 
+ \sum_{a = 1}^{4} R_{a} (\delta, \epsilon, \mu, \nu) \\
& = K (\nu, \mu) + K_{1}^{T} (\mu, \nu)\delta
+ K_{1}^{T} (\nu, \mu) \epsilon
\\ & \hspace{1cm}
+ \epsilon^{T}  K_{11} (\mu, \nu)\delta
+ {\frac{1}{2}} \delta^{T} K_{2} (\mu, \nu) \delta  
+ {\frac{1}{2}} \epsilon^{T} K_{2} (\nu, \mu) \epsilon 
+ \sum_{a = 1}^{4} R_{a} (\delta, \epsilon, \mu, \nu). 
\ean

Note now that 
\ban 
S_{i} & = {\frac {X_{i}}{D (X_{i})}} \cI_{\{ \cA_{i} \}} \\
& = \Delta^{-1} \bigl( \mu + \Sigma^{1/2} Z_{i} \bigr) 
- \Delta^{-1} \bigl( \mu + \Sigma^{1/2} Z_{i} \bigr) \cI_{\{ \cA_{i}^{C} \}}
\\ & \hspace{1cm}
- \Delta^{-2} \bigl( \mu + \Sigma^{1/2} Z_{i} \bigr) 
\bigl( D(X_{i}) - \Delta \bigr) \bigl( 1 - {\frac{R_{i}}{1 + R_{i}}} \bigr)
\cI_{\{ \cA_{i} \}} \\
& = \Delta^{-1} \bigl( \mu + \Sigma^{1/2} Z_{i}  - U_{i} \bigr). 
\ean

Thus  we have
\ban 
&  K \bigl( S_{i},  S_{j} \bigr) 
 = K (\Delta^{-1} \mu, \Delta^{-1} \mu) 
\\ & \hspace{1cm}
+ K_{1}^{T} (\Delta^{-1} \mu, \Delta^{-1} \mu)\Delta^{-1} \bigl(\Sigma^{1/2} Z_{i}  - U_{i} \bigr)
+ K_{1}^{T} (\Delta^{-1} \mu, \Delta^{-1} \mu) \Delta^{-1} \bigl(\Sigma^{1/2} Z_{j}  - U_{j} \bigr)
\\ & \hspace{1cm}
+ \Delta^{-1} \bigl(\Sigma^{1/2} Z_{j}  - U_{j} \bigr)^{T}  K_{11} (\Delta^{-1} \mu, \Delta^{-1} \mu)\Delta^{-1} \bigl(\Sigma^{1/2} Z_{i}  - U_{i} \bigr)
\\ & \hspace{1cm}
+ {\frac{1}{2}} \Delta^{-1} \bigl(\Sigma^{1/2} Z_{i}  - U_{i} \bigr)^{T} K_{2} (\Delta^{-1} \mu, \Delta^{-1} \mu) \Delta^{-1} \bigl(\Sigma^{1/2} Z_{i}  - U_{i} \bigr) 
\\ & \hspace{1cm}
+ {\frac{1}{2}} \Delta^{-1} \bigl(\Sigma^{1/2} Z_{j}  - U_{j} \bigr)^{T} K_{2} (\Delta^{-1} \mu, \Delta^{-1} \mu) \Delta^{-1} \bigl(\Sigma^{1/2} Z_{j}  - U_{j} \bigr)
\\ & \hspace{1cm}
+ \sum_{a = 1}^{4} R_{a} \bigl( \Delta^{-1} \bigl(\Sigma^{1/2} Z_{i}  - U_{i} \bigr), \Delta^{-1} \bigl(\Sigma^{1/2} Z_{j}  - U_{j} \bigr), \Delta^{-1} \mu, \Delta^{-1} \mu \bigr)\\
&  = K (\Delta^{-1} \mu, \Delta^{-1} \mu) 
+ \Delta^{-1}  K_{1}^{T} (\Delta^{-1} \mu, \Delta^{-1} \mu)\Sigma^{1/2} Z_{i} 
+ \Delta^{-1} K_{1}^{T} (\Delta^{-1} \mu, \Delta^{-1} \mu)  \Sigma^{1/2} Z_{j} 
\\ & \hspace{1cm}
+ \Delta^{-2}  Z_{j}^{T} \Sigma^{1/2} K_{11} (\Delta^{-1} \mu, \Delta^{-1} \mu) \Sigma^{1/2} Z_{i}
\\ & \hspace{1cm}
+ {\frac{1}{2}} \Delta^{-2} Z_{i}^{T} \Sigma^{1/2} K_{2} (\Delta^{-1} \mu, \Delta^{-1} \mu) \Sigma^{1/2} Z_{i} 
\\ & \hspace{1cm}
+ {\frac{1}{2}} \Delta^{-2} Z_{j}^{T} \Sigma^{1/2} K_{2} (\Delta^{-1} \mu, \Delta^{-1} \mu) \Delta^{-1} \Sigma^{1/2} Z_{j} 
\\ & \hspace{1cm}
+ \sum_{a = 1}^{15} R_{a} \bigl( \Delta^{-1} \bigl(\Sigma^{1/2} Z_{i}  - U_{i} \bigr), \Delta^{-1} \bigl(\Sigma^{1/2} Z_{j}  - U_{j} \bigr), \Delta^{-1} \mu, \Delta^{-1} \mu \bigr). 
\ean

In the above, $R_{a}$ are small order (and hence negligible) terms, for all 
choices of $a$. The different cases then involve detailed analysis of the leading terms. We omit the routine algebraic details here. 
\end{proof}

\begin{proof}[Proof of Theorem 3]
With some initial specifications, this proof follows steps of the proof of proposition 9 in \cite{rustamov2021intrinsic}. We define a new random variable $R$ as
$$ R = \left( \frac{1}{\rho_1} + \frac{1}{\rho_2} \right)^{-1}
\left[ \frac{1}{\rho_1} (X_1 - \mu_1) + \frac{1}{\rho_2} (X_2 - \mu_2) \right]
= \rho_2 ( X_1 - \mu_1) + \rho_1 ( X_2 - \mu_2) .$$
 Suppose $\gamma_{k}, \psi_k ,k=1,2,\ldots$ are the eigenvalues and eigenfunctions of 
$$
\frac{1}{\rho_{1}\rho_{2}}\int K( S(x),S(x'))
\psi_{k}(x')dR(x')=\gamma_{k}\psi_{k}(x).
$$
Then we shall prove that under $H_{0}^2: \mu_1 = \mu_2$,
$$ a_p^2 n T_{2,K,n_1,n_2} \leadsto \sum_{k=1}^\infty \gamma_k (A_k-1),$$
where $A_{k}$ are i.i.d. $\chi_1^2$ random variables. Under
$H_{a}^2: \mu_1 \neq \mu_2$, we have $a_p \sqrt{n}(T_{2,K,n_1,n_2} - T_2)\leadsto N(0,\sigma_{1}^{2})$,
where 
\begin{eqnarray*}
\sigma_{1}^{2} &= & 4\left[
\frac{1}{\rho_{1}} \mathbb{V}_{P \sim F_1} \mathbb{E}_{P' \sim F_1} K(S(P),S(P')) +\frac{1}{\rho_{2}} \mathbb{V}_{Q \sim F_2} \mathbb{E}_{Q' \sim F_2} K(S(Q),S(Q'))
\right.\nonumber \\
&  & + \left.\frac{1}{\rho_{1}} \mathbb{V}_{P \sim F_1} \mathbb{E}_{Q \sim F_2} K(S(P),S(Q)) +\frac{1}{\rho_{2}} \mathbb{V}_{Q \sim F_2} \mathbb{E}_{P \sim F_1} K(S(P),S(Q))
\right],
\end{eqnarray*}
where we denote by $F_1, F_2$ the distributions of $Y_1, Y_2$, respectively. aFollowing this, the steps of the proof follow along the lines of that in Proposition 9 in \cite{rustamov2021intrinsic}, page 28 onwards. Specifically, the Hilbert inner products and Hilbert centroids therein are replaced by the kernel and population mean parameters, respectively, while The multiplier $N$ gets replaced by $a_p^2 n$.
\end{proof}

\begin{proof}[Proof of Corollary 1]
This proof follows the exact same steps as that of Theorem 3 in \cite{rustamov2021intrinsic}, with the same notation replacements as above.
\end{proof}
\section{Additional numerical results}
\label{sec:num}

\subsection{Synthetic data}
We present the complete set of results on all our simulation settings in tables \ref{tab:supptable1} and \ref{tab:supptable2}. Generalized sign-based tests consistently demonstrate better power across all settings, as compared to the CQ and WPL tests. The $\ell_1$ and $\ell_\infty$ norm based tests perform comparatively better in sparse and dense mean settings, respectively. Interestingly, For the two-sample scenarios, these respective norm combinations also maintain the nominal size.

\begin{sidewaystable}
\centering
\scalebox{.8}{
\begin{tabular}{ccccccccccccccccccc}
\toprule
\multicolumn{18}{c}{Setting 1: AR $\Sigma$, dense $\mu$, $t_3$}\\\midrule
$\delta$ &
$(\ell_1,1,1)$ & $(\ell_1,4,4)$ & \boldmath $(\ell_1,4,p)$ \unboldmath & $(\ell_1,p,4)$ & $(\ell_1,p,p)$ &
$(\ell_2,1,1)$ & \boldmath $(\ell_2,4,4)$ \unboldmath & $(\ell_2,4,p)$ & $(\ell_2,p,4)$ & \boldmath $(\ell_2,p,p)$ \unboldmath &
$(\ell_\infty,1,1)$ & $(\ell_\infty,4,4)$ & $(\ell_\infty,4,p)$ & $(\ell_\infty,p,4)$ & $(\ell_\infty,p,p)$ & CQ & WPL \\\midrule
0.00 & \textbf{0.05} & \textbf{0.05} & \textbf{0.05} & \textbf{0.05} & \textbf{0.05} & \textbf{0.05} & \textbf{0.05} & \textbf{0.06} & \textbf{0.05} & \textbf{0.05} & \textbf{0.05} & \textbf{0.04} & \textbf{0.03} & \textbf{0.05} & 0.06 & \textbf{0.05} & \textbf{0.05} \\ 
  0.02 & 0.09 & 0.09 & 0.09 & 0.09 & 0.09 & 0.09 & 0.09 & 0.07 & 0.09 & 0.09 & 0.09 & 0.08 & 0.03 & 0.09 & 0.07 & 0.06 & 0.08 \\ 
  0.04 & 0.36 & 0.36 & 0.37 & 0.36 & 0.36 & 0.36 & 0.37 & 0.09 & 0.36 & 0.37 & 0.36 & 0.33 & 0.04 & 0.36 & 0.13 & 0.12 & 0.29 \\ 
  0.06 & 0.82 & 0.82 & 0.82 & 0.82 & 0.82 & 0.82 & 0.82 & 0.14 & 0.82 & 0.82 & 0.81 & 0.76 & 0.05 & 0.81 & 0.28 & 0.31 & 0.76 \\ 
  0.08 & 1.00 & 1.00 & 1.00 & 1.00 & 1.00 & 1.00 & 1.00 & 0.25 & 1.00 & 1.00 & 1.00 & 0.98 & 0.08 & 1.00 & 0.48 & 0.62 & 0.99 \\ 
  0.10 & 1.00 & 1.00 & 1.00 & 1.00 & 1.00 & 1.00 & 1.00 & 0.39 & 1.00 & 1.00 & 1.00 & 1.00 & 0.12 & 1.00 & 0.68 & 0.86 & 1.00 \\ 
  0.12 & 1.00 & 1.00 & 1.00 & 1.00 & 1.00 & 1.00 & 1.00 & 0.55 & 1.00 & 1.00 & 1.00 & 1.00 & 0.19 & 1.00 & 0.85 & 0.96 & 1.00 \\
\midrule
\multicolumn{18}{c}{Setting 2: AR $\Sigma$, sparse $\mu$, $t_3$}\\\midrule
$\delta$ &
\boldmath $(\ell_1,1,1)$ \unboldmath & \boldmath $(\ell_1,4,4)$ \unboldmath & \boldmath $(\ell_1,4,p)$ \unboldmath & \boldmath $(\ell_1,p,4)$ \unboldmath & \boldmath $(\ell_1,p,p)$ \unboldmath &
\boldmath $(\ell_2,1,1)$ & $(\ell_2,4,4)$ & $(\ell_2,4,p)$ & \boldmath $(\ell_2,p,4)$ \unboldmath & \boldmath $(\ell_2,p,p)$ \unboldmath &
$(\ell_\infty,1,1)$ & $(\ell_\infty,4,4)$ & $(\ell_\infty,4,p)$ & $(\ell_\infty,p,4)$ & $(\ell_\infty,p,p)$ & CQ & WPL \\\midrule 
0.00 & \textbf{0.05} & \textbf{0.05} & \textbf{0.05} & \textbf{0.05} & \textbf{0.05} & \textbf{0.05} &\textbf{ 0.05} & 0.06 & \textbf{0.05} & \textbf{0.05} & \textbf{0.05} & \textbf{0.04} & \textbf{0.03} & \textbf{0.05} & 0.06 & \textbf{0.05} & \textbf{0.05} \\ 
  0.15 & 0.08 & 0.08 & 0.08 & 0.08 & 0.08 & 0.08 & 0.08 & 0.06 & 0.08 & 0.08 & 0.08 & 0.06 & 0.02 & 0.08 & 0.07 & 0.05 & 0.06 \\ 
  0.30 & 0.20 & 0.20 & 0.20 & 0.20 & 0.20 & 0.20 & 0.20 & 0.07 & 0.20 & 0.20 & 0.19 & 0.18 & 0.02 & 0.19 & 0.09 & 0.08 & 0.14 \\ 
  0.45 & 0.56 & 0.56 & 0.56 & 0.56 & 0.56 & 0.56 & 0.56 & 0.10 & 0.56 & 0.56 & 0.56 & 0.49 & 0.03 & 0.56 & 0.17 & 0.16 & 0.48 \\ 
  0.60 & 0.90 & 0.90 & 0.90 & 0.90 & 0.90 & 0.90 & 0.90 & 0.15 & 0.90 & 0.90 & 0.88 & 0.84 & 0.05 & 0.88 & 0.28 & 0.37 & 0.86 \\ 
  0.75 & 1.00 & 1.00 & 1.00 & 1.00 & 1.00 & 1.00 & 1.00 & 0.23 & 1.00 & 1.00 & 1.00 & 0.98 & 0.06 & 1.00 & 0.43 & 0.64 & 0.99 \\ 
  0.90 & 1.00 & 1.00 & 1.00 & 1.00 & 1.00 & 1.00 & 1.00 & 0.34 & 1.00 & 1.00 & 1.00 & 1.00 & 0.08 & 1.00 & 0.59 & 0.84 & 1.00 \\ 
\midrule
\multicolumn{18}{c}{Setting 3: SAR $\Sigma$, dense $\mu$, $t_3$}\\\midrule
$\delta$ &
$(\ell_1,1,1)$ & $(\ell_1,4,4)$ & $(\ell_1,4,p)$ & $(\ell_1,p,4)$ & $(\ell_1,p,p)$ &
$(\ell_2,1,1)$ & $(\ell_2,4,4)$ & $(\ell_2,4,p)$ & $(\ell_2,p,4)$ & $(\ell_2,p,p)$ &
$(\ell_\infty,1,1)$ & $(\ell_\infty,4,4)$ & $(\ell_\infty,4,p)$ & $(\ell_\infty,p,4)$ & \boldmath $(\ell_\infty,p,p)$ \unboldmath & CQ & WPL \\\midrule 
0.00 & \textbf{0.04} & \textbf{0.04} & \textbf{0.04} & \textbf{0.04} & \textbf{0.04} & \textbf{0.03} & \textbf{0.03} & \textbf{0.04} & \textbf{0.03} & \textbf{0.03} & \textbf{0.03} & \textbf{0.03} & \textbf{0.04} & \textbf{0.03} & \textbf{0.03} & \textbf{0.04} & \textbf{0.04} \\ 
  0.10 & 0.06 & 0.06 & 0.06 & 0.06 & 0.06 & 0.06 & 0.06 & 0.05 & 0.06 & 0.06 & 0.06 & 0.06 & 0.05 & 0.06 & 0.06 & 0.04 & 0.05 \\ 
  0.20 & 0.21 & 0.21 & 0.20 & 0.21 & 0.21 & 0.27 & 0.27 & 0.09 & 0.27 & 0.27 & 0.30 & 0.31 & 0.08 & 0.30 & 0.32 & 0.08 & 0.26 \\ 
  0.30 & 0.71 & 0.71 & 0.68 & 0.71 & 0.71 & 0.91 & 0.91 & 0.21 & 0.91 & 0.92 & 0.95 & 0.96 & 0.16 & 0.95 & 0.97 & 0.19 & 0.90 \\ 
  0.40 & 1.00 & 1.00 & 1.00 & 1.00 & 1.00 & 1.00 & 1.00 & 0.37 & 1.00 & 1.00 & 1.00 & 1.00 & 0.33 & 1.00 & 1.00 & 0.50 & 1.00 \\ 
  0.50 & 1.00 & 1.00 & 1.00 & 1.00 & 1.00 & 1.00 & 1.00 & 0.57 & 1.00 & 1.00 & 1.00 & 1.00 & 0.55 & 1.00 & 1.00 & 0.84 & 1.00 \\ 
  0.60 & 1.00 & 1.00 & 1.00 & 1.00 & 1.00 & 1.00 & 1.00 & 0.74 & 1.00 & 1.00 & 1.00 & 1.00 & 0.66 & 1.00 & 1.00 & 0.96 & 1.00 \\ 
\midrule
\multicolumn{18}{c}{Setting 4: SAR $\Sigma$, sparse $\mu$, $t_3$}\\\midrule
$\delta$ &
\boldmath $(\ell_1,1,1)$ \unboldmath & \boldmath $(\ell_1,4,4)$ \unboldmath & $(\ell_1,4,p)$ & \boldmath $(\ell_1,p,4)$ \unboldmath & \boldmath $(\ell_1,p,p)$ \unboldmath &
$(\ell_2,1,1)$ & $(\ell_2,4,4)$ & $(\ell_2,4,p)$ & $(\ell_2,p,4)$ & $(\ell_2,p,p)$ &
$(\ell_\infty,1,1)$ & $(\ell_\infty,4,4)$ & $(\ell_\infty,4,p)$ & $(\ell_\infty,p,4)$ & $(\ell_\infty,p,p)$ & CQ & WPL \\\midrule 
0.00 & \textbf{0.04} & \textbf{0.04} & \textbf{0.04} & \textbf{0.04} & \textbf{0.04} & \textbf{0.03} & \textbf{0.03} & \textbf{0.04} & \textbf{0.03} & \textbf{0.03} & \textbf{0.03} & \textbf{0.03} & \textbf{0.04} & \textbf{0.03} & \textbf{0.03} & \textbf{0.04} & \textbf{0.04} \\ 
  1.05 & 0.10 & 0.10 & 0.10 & 0.10 & 0.10 & 0.09 & 0.09 & 0.04 & 0.09 & 0.09 & 0.09 & 0.09 & 0.04 & 0.09 & 0.09 & 0.07 & 0.09 \\ 
  2.10 & 0.31 & 0.31 & 0.30 & 0.31 & 0.31 & 0.28 & 0.28 & 0.07 & 0.28 & 0.28 & 0.28 & 0.28 & 0.05 & 0.28 & 0.27 & 0.14 & 0.28 \\ 
  3.15 & 0.65 & 0.65 & 0.65 & 0.65 & 0.65 & 0.62 & 0.61 & 0.12 & 0.62 & 0.62 & 0.58 & 0.58 & 0.07 & 0.58 & 0.57 & 0.33 & 0.61 \\ 
  4.20 & 0.90 & 0.90 & 0.90 & 0.90 & 0.90 & 0.88 & 0.88 & 0.18 & 0.88 & 0.88 & 0.86 & 0.86 & 0.10 & 0.86 & 0.84 & 0.57 & 0.88 \\ 
  5.25 & 0.99 & 0.99 & 0.99 & 0.99 & 0.99 & 0.98 & 0.98 & 0.29 & 0.98 & 0.98 & 0.97 & 0.97 & 0.14 & 0.97 & 0.97 & 0.77 & 0.98 \\ 
  6.30 & 1.00 & 1.00 & 1.00 & 1.00 & 1.00 & 1.00 & 1.00 & 0.44 & 1.00 & 1.00 & 1.00 & 1.00 & 0.19 & 1.00 & 1.00 & 0.89 & 1.00 \\ 
\bottomrule
\end{tabular}
}
\caption{Performance comparisons for one-sample tests. Best performing tests are marked in bold. At $\delta=0$, all tests with size $\leq 0.05$ are marked in bold.}
\label{tab:supptable1}
\end{sidewaystable}

\begin{sidewaystable}
\centering
\scalebox{.8}{
\begin{tabular}{ccccccccccccccccccc}
\toprule
\multicolumn{18}{c}{Setting 1: AR $\Sigma$, dense $\mu_2$, MVG}\\\midrule
$\delta$ &
\boldmath $(\ell_1,1,1)$ \unboldmath & \boldmath $(\ell_1,4,4)$ \unboldmath & \boldmath $(\ell_1,4,p)$ \unboldmath & \boldmath $(\ell_1,p,4)$ \unboldmath & \boldmath $(\ell_1,p,p)$ \unboldmath &
\boldmath $(\ell_2,1,1)$ \unboldmath & $(\ell_2,4,4)$ & $(\ell_2,4,p)$ & $(\ell_2,p,4)$ & $(\ell_2,p,p)$ &
$(\ell_\infty,1,1)$ & $(\ell_\infty,4,4)$ & $(\ell_\infty,4,p)$ & $(\ell_\infty,p,4)$ & $(\ell_\infty,p,p)$ & CQ & WPL \\\midrule
0.00 & \textbf{0.05} & \textbf{0.05} & \textbf{0.05} & \textbf{0.05} & \textbf{0.05} & \textbf{0.05} & \textbf{0.05} & \textbf{0.04} & \textbf{0.05} & \textbf{0.04} & \textbf{0.05} & 0.06 & 0.06 & 0.06 & 0.06 & \textbf{0.05} & \textbf{0.05} \\ 
  0.03 & 0.09 & 0.09 & 0.08 & 0.09 & 0.08 & 0.09 & 0.08 & 0.04 & 0.08 & 0.04 & 0.08 & 0.07 & 0.07 & 0.07 & 0.07 & 0.07 & 0.07 \\ 
  0.06 & 0.22 & 0.22 & 0.22 & 0.22 & 0.22 & 0.22 & 0.23 & 0.06 & 0.23 & 0.06 & 0.22 & 0.11 & 0.11 & 0.11 & 0.11 & 0.16 & 0.17 \\ 
  0.09 & 0.56 & 0.56 & 0.56 & 0.56 & 0.56 & 0.56 & 0.56 & 0.07 & 0.56 & 0.07 & 0.54 & 0.21 & 0.21 & 0.21 & 0.21 & 0.48 & 0.48 \\ 
  0.12 & 0.90 & 0.90 & 0.90 & 0.90 & 0.90 & 0.90 & 0.89 & 0.12 & 0.89 & 0.12 & 0.89 & 0.44 & 0.44 & 0.44 & 0.44 & 0.86 & 0.86 \\ 
  0.15 & 1.00 & 1.00 & 1.00 & 1.00 & 1.00 & 1.00 & 0.99 & 0.21 & 0.99 & 0.21 & 0.99 & 0.75 & 0.75 & 0.75 & 0.75 & 0.99 & 0.99 \\ 
  0.18 & 1.00 & 1.00 & 1.00 & 1.00 & 1.00 & 1.00 & 1.00 & 0.35 & 1.00 & 0.35 & 1.00 & 0.95 & 0.95 & 0.95 & 0.95 & 1.00 & 1.00 \\ 
\midrule
\multicolumn{18}{c}{Setting 2: AR $\Sigma$, sparse $\mu_2$, MVG}\\\midrule
$\delta$ &
\boldmath $(\ell_1,1,1)$ \unboldmath & \boldmath $(\ell_1,4,4)$ \unboldmath & $(\ell_1,4,p)$ & \boldmath $(\ell_1,p,4)$ \unboldmath & $(\ell_1,p,p)$ &
$(\ell_2,1,1)$ & $(\ell_2,4,4)$ & $(\ell_2,4,p)$ & $(\ell_2,p,4)$ & $(\ell_2,p,p)$ &
$(\ell_\infty,1,1)$ & $(\ell_\infty,4,4)$ & $(\ell_\infty,4,p)$ & $(\ell_\infty,p,4)$ & $(\ell_\infty,p,p)$ & CQ & WPL \\\midrule
0.00 & 0.07 & 0.07 & 0.06 & 0.07 & 0.06 & 0.06 & 0.07 & 0.06 & 0.07 & 0.06 & 0.06 & \textbf{0.05} & \textbf{0.04} & \textbf{0.05} & \textbf{0.04} & 0.06 & 0.07 \\ 
  3.00 & 0.20 & 0.20 & 0.13 & 0.20 & 0.13 & 0.18 & 0.19 & 0.06 & 0.19 & 0.06 & 0.16 & 0.12 & 0.04 & 0.12 & 0.04 & 0.19 & 0.18 \\ 
  6.00 & 0.67 & 0.67 & 0.34 & 0.67 & 0.34 & 0.63 & 0.64 & 0.12 & 0.64 & 0.12 & 0.62 & 0.48 & 0.08 & 0.48 & 0.08 & 0.67 & 0.64 \\ 
  9.00 & 0.97 & 0.97 & 0.69 & 0.97 & 0.69 & 0.94 & 0.95 & 0.26 & 0.95 & 0.26 & 0.93 & 0.86 & 0.10 & 0.86 & 0.10 & 0.96 & 0.95 \\ 
  12.00 & 1.00 & 1.00 & 0.90 & 1.00 & 0.90 & 1.00 & 1.00 & 0.45 & 1.00 & 0.45 & 1.00 & 0.98 & 0.15 & 0.98 & 0.15 & 1.00 & 1.00 \\ 
  15.00 & 1.00 & 1.00 & 0.97 & 1.00 & 0.97 & 1.00 & 1.00 & 0.65 & 1.00 & 0.65 & 1.00 & 1.00 & 0.17 & 1.00 & 0.17 & 1.00 & 1.00 \\ 
  18.00 & 1.00 & 1.00 & 0.99 & 1.00 & 0.99 & 1.00 & 1.00 & 0.82 & 1.00 & 0.82 & 1.00 & 1.00 & 0.22 & 1.00 & 0.22 & 1.00 & 1.00 \\ 
\midrule
\multicolumn{18}{c}{Setting 3: SAR $\Sigma$, dense $\mu_2$, MVG}\\\midrule
$\delta$ &
$(\ell_1,1,1)$ & $(\ell_1,4,4)$ & $(\ell_1,4,p)$ & $(\ell_1,p,4)$ & $(\ell_1,p,p)$ &
$(\ell_2,1,1)$ & \boldmath $(\ell_2,4,4)$ \unboldmath & $(\ell_2,4,p)$ & \boldmath $(\ell_2,p,4)$ \unboldmath & $(\ell_2,p,p)$ &
$(\ell_\infty,1,1)$ & $(\ell_\infty,4,4)$ & $(\ell_\infty,4,p)$ & $(\ell_\infty,p,4)$ & $(\ell_\infty,p,p)$ & CQ & WPL \\\midrule
0.00 & \textbf{0.05} & \textbf{0.05} & \textbf{0.05} & \textbf{0.05} & \textbf{0.05} & \textbf{0.05} & \textbf{0.05} & \textbf{0.04} & \textbf{0.05} & \textbf{0.04} & \textbf{0.05} & 0.06 & 0.06 & 0.06 & 0.06 & \textbf{0.05} & \textbf{0.05} \\ 
  0.03 & 0.09 & 0.09 & 0.08 & 0.09 & 0.08 & 0.09 & 0.08 & 0.04 & 0.08 & 0.04 & 0.08 & 0.07 & 0.07 & 0.07 & 0.07 & 0.07 & 0.07 \\ 
  0.06 & 0.22 & 0.22 & 0.22 & 0.22 & 0.22 & 0.22 & 0.23 & 0.06 & 0.23 & 0.06 & 0.22 & 0.11 & 0.11 & 0.11 & 0.11 & 0.16 & 0.17 \\ 
  0.09 & 0.56 & 0.56 & 0.56 & 0.56 & 0.56 & 0.56 & 0.56 & 0.07 & 0.56 & 0.07 & 0.54 & 0.21 & 0.21 & 0.21 & 0.21 & 0.48 & 0.48 \\ 
  0.12 & 0.90 & 0.90 & 0.90 & 0.90 & 0.90 & 0.90 & 0.89 & 0.12 & 0.89 & 0.12 & 0.89 & 0.44 & 0.44 & 0.44 & 0.44 & 0.86 & 0.86 \\ 
  0.15 & 1.00 & 1.00 & 1.00 & 1.00 & 1.00 & 1.00 & 0.99 & 0.21 & 0.99 & 0.21 & 0.99 & 0.75 & 0.75 & 0.75 & 0.75 & 0.99 & 0.99 \\ 
  0.18 & 1.00 & 1.00 & 1.00 & 1.00 & 1.00 & 1.00 & 1.00 & 0.35 & 1.00 & 0.35 & 1.00 & 0.95 & 0.95 & 0.95 & 0.95 & 1.00 & 1.00 \\
\midrule
\multicolumn{18}{c}{Setting 4: SAR $\Sigma$, sparse $\mu_2$, MVG}\\\midrule
$\delta$ &
\boldmath $(\ell_1,1,1)$ \unboldmath & \boldmath $(\ell_1,4,4)$ \unboldmath & $(\ell_1,4,p)$ & $(\ell_1,p,4)$ & \boldmath $(\ell_1,p,p)$ \unboldmath &
$(\ell_2,1,1)$ & $(\ell_2,4,4)$ & $(\ell_2,4,p)$ & $(\ell_2,p,4)$ & $(\ell_2,p,p)$ &
$(\ell_\infty,1,1)$ & $(\ell_\infty,4,4)$ & $(\ell_\infty,4,p)$ & $(\ell_\infty,p,4)$ & $(\ell_\infty,p,p)$ & CQ & WPL \\\midrule
0.00 & 0.07 & 0.07 & 0.06 & 0.07 & 0.06 & 0.06 & 0.07 & 0.06 & 0.07 & 0.06 & 0.06 & \textbf{0.05} & \textbf{0.04} & \textbf{0.05} & \textbf{0.04} & 0.06 & 0.07 \\ 
  3.00 & 0.20 & 0.20 & 0.13 & 0.20 & 0.13 & 0.18 & 0.19 & 0.06 & 0.19 & 0.06 & 0.16 & 0.12 & 0.04 & 0.12 & 0.04 & 0.19 & 0.18 \\ 
  6.00 & 0.67 & 0.67 & 0.34 & 0.67 & 0.34 & 0.63 & 0.64 & 0.12 & 0.64 & 0.12 & 0.62 & 0.48 & 0.08 & 0.48 & 0.08 & 0.67 & 0.64 \\ 
  9.00 & 0.97 & 0.97 & 0.69 & 0.97 & 0.69 & 0.94 & 0.95 & 0.26 & 0.95 & 0.26 & 0.93 & 0.86 & 0.10 & 0.86 & 0.10 & 0.96 & 0.95 \\ 
  12.00 & 1.00 & 1.00 & 0.90 & 1.00 & 0.90 & 1.00 & 1.00 & 0.45 & 1.00 & 0.45 & 1.00 & 0.98 & 0.15 & 0.98 & 0.15 & 1.00 & 1.00 \\ 
  14.00 & 1.00 & 1.00 & 0.96 & 1.00 & 0.96 & 1.00 & 1.00 & 0.58 & 1.00 & 0.58 & 1.00 & 1.00 & 0.16 & 1.00 & 0.16 & 1.00 & 1.00 \\ 
  16.00 & 1.00 & 1.00 & 0.98 & 1.00 & 0.98 & 1.00 & 1.00 & 0.71 & 1.00 & 0.71 & 1.00 & 1.00 & 0.18 & 1.00 & 0.18 & 1.00 & 1.00 \\  
\bottomrule
\end{tabular}
}
\caption{Performance comparisons for two-sample tests. Best performing tests are marked in bold. At $\delta=0$, all tests with size $\leq 0.05$ are marked in bold.}
\label{tab:supptable2}
\end{sidewaystable}

\subsection{MNIST digits}

\begin{figure}[t]
\centering
\begin{subfigure}[b]{0.3\linewidth}
\caption{CQ test}
\label{subfig:m1s}
\includegraphics[width=\linewidth]{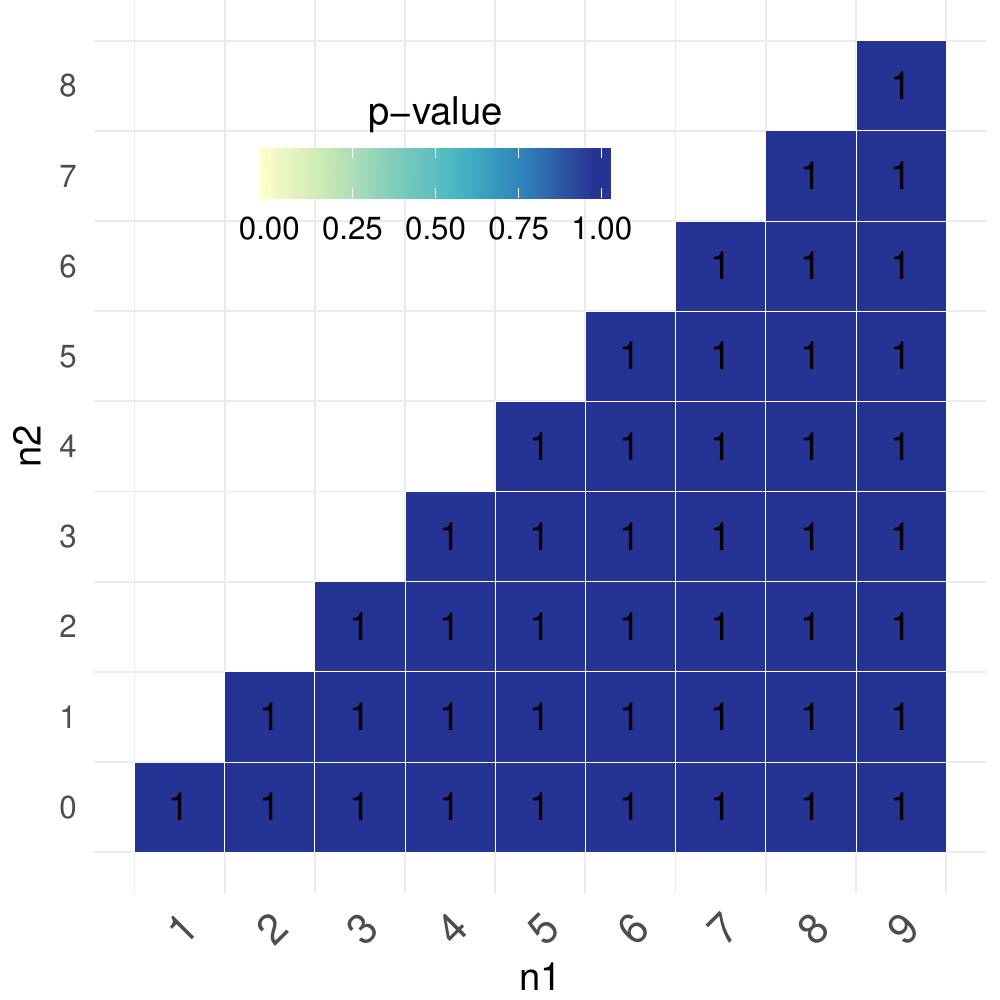}
\end{subfigure}
\begin{subfigure}[b]{0.3\linewidth}
\caption{WPL test}
\label{subfig:m2s}
\includegraphics[width=\textwidth]{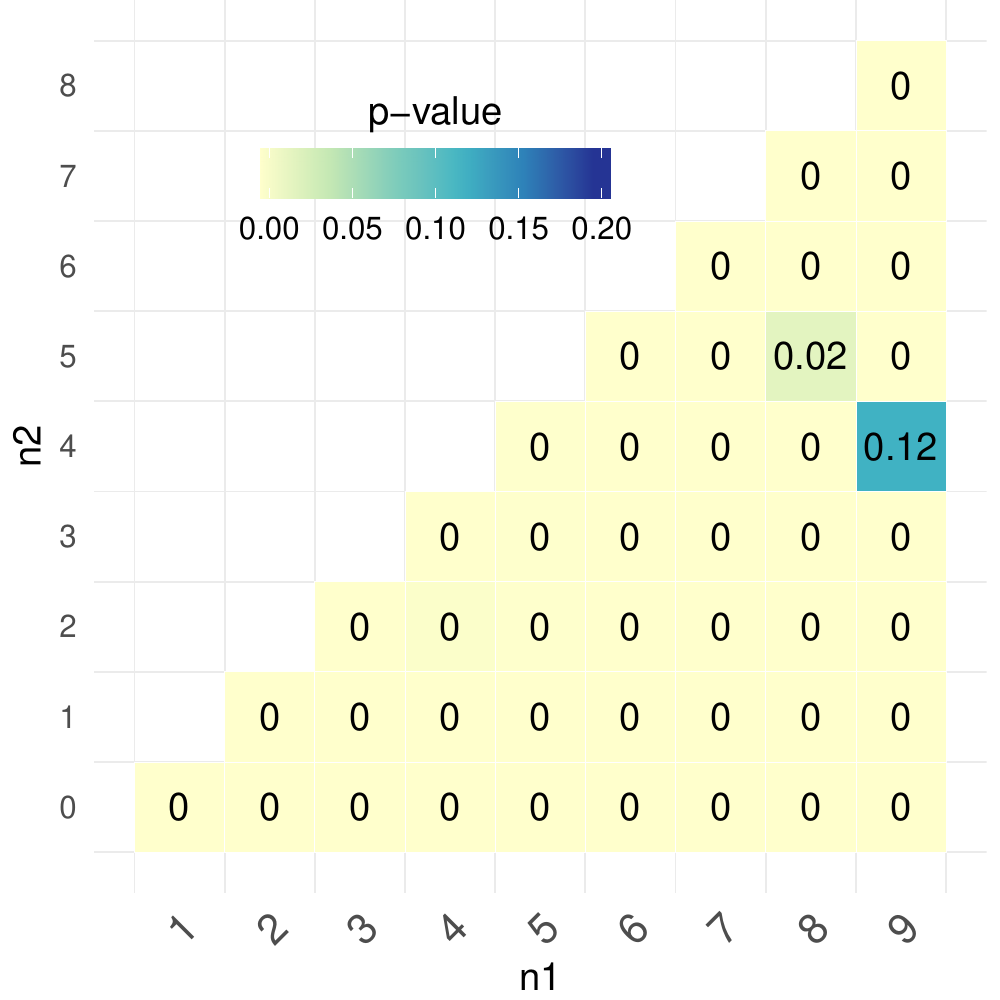}
\end{subfigure}
\begin{subfigure}[b]{0.3\linewidth}
\caption{$\ell_1$ norm, $a=1, b=1$}
\label{subfig:m3s}
\includegraphics[width=\linewidth]{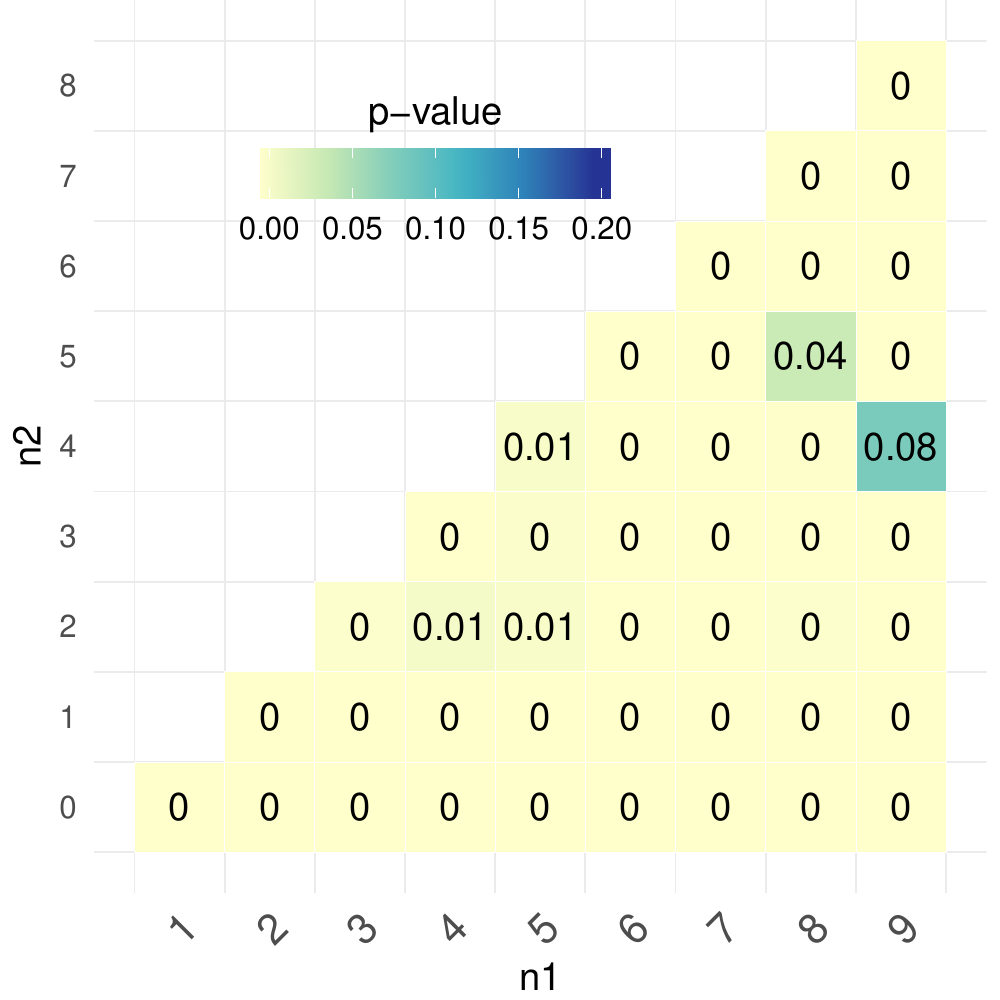}
\end{subfigure}

\caption{Outputs on MNIST dataset: $p$-values from pairwise comparison of digits: additional results.}
\label{fig:mnistsupp}
\end{figure}

Figure~\ref{fig:mnistsupp} presents results for additional tests on the pairwise MNIST digits. The CQ test \cite{ref:AoS10808_HDTest_ChenQin} based on untransformed high-dimensional samples performs poorly, and is not able to distinguish between any pair of digits. The WPL test~\cite{ref:JASA151658_HDTest_Lan} and the $\ell_1$ norm unkernelized tests perform similarly, and similar to the $\ell_2$ norm unkernelized test (see main paper, Fig.~3b), with the $\ell_1$ norm unkernelized test being marginally better than the other two.

\end{document}